\documentclass[final,1p,times,authoryear]{elsarticle}
\usepackage{amsmath}
\usepackage{amsthm}
\usepackage{amssymb}
\usepackage{amsfonts}

\newtheorem{theorem}{Theorem}
\newtheorem{lemma}[theorem]{Lemma}

\newtheorem{proposition}[theorem]{Proposition}

\newtheorem{example}[theorem]{Example}
\newtheorem{remark}[theorem]{Remark}

\usepackage[utf8]{inputenc}
\usepackage[T1]{fontenc}
\usepackage{graphicx}
\usepackage{hyperref}
\usepackage{longtable}
\usepackage{multirow}
\usepackage{float}
\usepackage{wrapfig}
\usepackage{rotating}
\usepackage[normalem]{ulem}
\usepackage{amsmath}
\usepackage{amsthm}
\usepackage{textcomp}
\usepackage{marvosym}
\usepackage{wasysym}
\usepackage{amssymb}
\usepackage{capt-of}
\usepackage{bm}
\usepackage{comment}
\usepackage{color}
\usepackage{enumerate} 
\usepackage{myalgo}%markus
\usepackage{bm}

\newcommand{\coq}{\text{\sc Coq}}
\newcommand{\hol}{\text{\sc Hol-light}}

\newcommand{\mohab}[1]{{\color{green} Mohab: #1}}
\newcommand{\victor}[1]{{\color{red} Victor: #1}}

\newcommand{\newjsc}[1]{{{\color{black}#1}}}
\newcommand{\revise}[1]{{{\color{black}#1}}}
\providecommand{\alert}[1]{\textbf{#1}}
\def\A{\mathbf{A}}
\def\B{\mathbf{B}}
\def\I{\mathbf{I}}
\def\Sb{\mathbb{S}}
\def\K{S}
\def\X{\mathbf{X}}
\def\Y{\mathbf{Y}}
\def\bmx{\bm{x}}
\def\x{\bmx}
\def\bmy{\bm{y}}
\def\y{\bmy}
\def\p{\mathbf{p}}
\def\f{\mathbf{f}}
\def\q{\mathbf{q}}
\def\z{\mathbf{z}}
\def\M{\mathbf{M}}

\DeclareMathOperator{\polytope}{\mathcal{C}}

\newcommand{\R}{\mathbb{R}}
\newcommand{\Z}{\mathbb{Z}}
\newcommand{\Q}{\mathbb{Q}}
\newcommand{\N}{\mathbb{N}}
\newcommand{\red}[1]{\textbf{{\color{red}#1}}}
\DeclareMathOperator{\bigo}{\mathcal{O}}
\DeclareMathOperator{\sgn}{sgn}
\DeclareMathOperator{\trace}{Tr}
\DeclareMathOperator{\bigotilde}{\overset{\sim}{\mathcal{O}}}
\newcommand{\hasrealroots}{\texttt{has\_real\_roots}}
\newcommand{\hasrealrootsfun}[1]{\texttt{has\_real\_roots}(#1)}
\newcommand{\newtonpolytope}{\texttt{newton\_polytope}}
\newcommand{\newtonpolytopefun}[1]{\texttt{newton\_polytope}(#1)}
\newcommand{\sumtwosquares}{\texttt{sum\_two\_squares}}
\newcommand{\sumtwosquaresfun}[2]{\texttt{sum\_two\_squares}(#1,#2)}
\newcommand{\sumofsquares}{\texttt{sos}}
\newcommand{\sumofsquaresfun}[2]{\texttt{sos}(#1,#2)}
\newcommand{\sdp}{\texttt{sdp}}
\newcommand{\sdpcon}{\texttt{sdp}}
\newcommand{\sdpHA}{\texttt{sdp}}
\newcommand{\roundfun}[2]{\texttt{round}(#1,#2)}
\newcommand{\ldlfun}[1]{\texttt{ldl}(#1)}
\newcommand{\sdpfun}[3]{\texttt{sdp}(#1,#2,#3)}
\newcommand{\sdpfunHA}[4]{\texttt{sdp}(#1,#2,#3,#4)}
\newcommand{\sdpconfun}[4]{\texttt{sdp}(#1,#2,#3,#4)}
\newcommand{\cholesky}{\texttt{cholesky}}
\newcommand{\choleskyfun}[3]{\texttt{\cholesky}(#1,#2,#3)}
\newcommand{\soslist}{\texttt{sos\_list}}
\newcommand{\nsdp}{n_\text{sdp}}
\newcommand{\msdp}{m_\text{sdp}}
\newcommand{\qlist}{\texttt{q\_list}}
\newcommand{\hlist}{\texttt{h\_list}}
\newcommand{\clist}{\texttt{c\_list}}
\newcommand{\calpha}{\texttt{c\_alpha}}
\newcommand{\slist}{\texttt{s\_list}}
\newcommand{\univsos}{\texttt{univsos}}
\newcommand{\univsosone}{\texttt{univsos1}}
\newcommand{\univsostwo}{\texttt{univsos2}}
\newcommand{\intsos}{\texttt{intsos}}
\newcommand{\multivsos}{\texttt{multivsos}}
\newcommand{\realcertify}{\texttt{RealCertify}}
\newcommand{\intsosfun}[5]{\texttt{intsos}(#1,#2,#3,#4,#5)}
\newcommand{\absorb}{\texttt{absorb}}
\newcommand{\absorbfun}[5]{\texttt{absorb}(#1,#2,#3,#4,#5)}
\newcommand{\putinarsosfun}[6]{\texttt{Putinarsos}(#1,#2,#3,#4,#5,#6)}
\newcommand{\reznicksos}{\texttt{Reznicksos}}
\newcommand{\interiorsoscone}{\texttt{interiorSOScone}}
\newcommand{\hilbertsos}{\texttt{Hilbertsos}}
\newcommand{\hilbertsosfun}[5]{\texttt{Hilbertsos}(#1,#2,#3,#4,#5)}
\newcommand{\cad}{\texttt{CAD}}
\newcommand{\PP}{\texttt{RoundProject}}
\newcommand{\PPfun}[4]{\texttt{RoundProject}(#1,#2,#3,#4)}
\newcommand{\PPcon}{\texttt{RoundProjectPutinar}}
\newcommand{\PPconfun}[7]{\texttt{RoundProjectPutinar}(#1,#2,#3,#4,#5,#6,#7)}
\newcommand{\raglib}{\texttt{RAGLib}}
\newcommand{\putinarsos}{\texttt{Putinarsos}}
\newcommand{\multivsosone}{\texttt{multivsos1}}
\newcommand{\multivsostwo}{\texttt{multivsos2}}

\theoremstyle{plain}
\newtheorem{assumption}[theorem]{Assumption}

\newcommand{\anon}[0]{\,\centerdot\,}
\newcommand{\spt}[1]{\mbox{supp}(#1)}
\newcommand{\bracket}[2]{\langle #1,#2\rangle}

\DeclareMathOperator{\vol}{vol}

\def\mohab#1{\textcolor{magenta}{#1}}
\definecolor{dkviolet}{rgb}{0.6,0,0.8}
\def\victor#1{\textcolor{dkviolet}{#1}}

\begin{document}

\begin{frontmatter}

\title{On Exact {Reznick}, Hilbert-Artin and Putinar's Representations}

\author{Victor Magron}
\address{CNRS; LAAS; 7 avenue du colonel Roche, F-31400 Toulouse; France}
\ead{vmagron@laas.fr}
\ead[url]{https://homepages.laas.fr/vmagron}

\author{Mohab Safey El Din}
\address{Sorbonne Universit\'e, CNRS, Laboratoire d'Informatique de Paris 6 (UMR
  7606), PolSys, Paris, France.}
\ead{mohab.safey@lip6.fr}
\ead[url]{http://www-polsys.lip6.fr/~safey}

\begin{abstract}
We consider the problem of {computing} exact sums of squares (SOS)
decompositions for certain classes of non-negative multivariate
polynomials, relying on semidefinite programming (SDP) solvers.

We provide a hybrid numeric-symbolic algorithm computing {\em exact} rational
SOS decompositions with rational coefficients for polynomials lying in the
interior of the SOS cone. The first step of this algorithm computes an
approximate SOS decomposition for a perturbation of the input polynomial with an
arbitrary-precision SDP solver. Next, an exact SOS decomposition is obtained thanks to
the perturbation terms and a compensation phenomenon. 
We prove that bit complexity estimates on output size and runtime are both singly exponential \revise{in the cardinality of the Newton polytope (or doubly exponential in the number of variables)}. 
Next, we apply this algorithm to compute exact {Reznick},
Hilbert-Artin's representation and Putinar's representations respectively for
positive definite forms and positive polynomials over basic compact
semi-algebraic sets. We also report on practical experiments done with the
implementation of these algorithms and existing alternatives such as the
critical point method and cylindrical algebraic decomposition.
\end{abstract}

\end{frontmatter}

%!TEX root = multivsosJSC.tex
\section{Introduction}
\label{sec:intro}

Let $\Q$ (resp.~$\R$) be the field of rational (resp.~real) numbers
and $X = (X_1, \ldots, X_n)$ be a sequence of variables. We consider
the problem of deciding the non-negativity of $f \in \Q[X]$ either
over $\R^n$ or over a {closed} semi-algebraic set $S$ defined by some
constraints $g_1\geq 0, \ldots, g_m\geq 0$ (with $g_j \in \Q[X]$).
Further, $d$ denotes the maximum of the total degrees of these
polynomials.

This problem is known to be co-NP hard~\citep{blum}. The Cylindrical Algebraic
Decomposition algorithm due to~\cite{Collins75} {and \cite{Wut76}} allows one to
solve it in time doubly exponential in $n$ (and polynomial in $d$). This has
been significantly improved, through the so-called critical point method,
starting from~\cite{GV88} which culminates with~\cite{BPR98} to establish that
this decision problem can be solved in time $((m+1)d)^{O(n)}$. These latter ones
have been developed to obtain practically fast implementations which reflect the
complexity gain (see
%e.g.~\cite{BGHM1,BGHM2,BGHM3,SaSc03,S07,BGHSS,BGHS14}) but still
e.g.~\cite{BGHM1,BGHM3,SaSc03,S07,BGHSS,
  Safey10Issac,BGHS14,Greuet14,Greuet12Sos}). These algorithms are ``root
finding'' ones: {they are designed to compute at least one point in each
connected component of the set defined by $f < 0$. This is done by solving
polynomial systems defining critical points of some well-chosen polynomial maps
restricted to $f=-\varepsilon$ for $\varepsilon$ small enough. Hence the
complexity of these algorithms depends on the difficulty of solving these
polynomial systems (which can be exponential in $n$ as the B\'ezout bound on the
number of their solutions is). Moreover, when $f$ is non-negative, they return an
empty list without a {\it certificate} that can be checked {\it a posteriori}.}
This paper focuses on the computation of such certificates under some favorable
situations.

To compute certificates of non-negativity, an approach based on {\it sums of
  squares} (SOS) decompositions of polynomials (see \cite{Las01sos} and
\cite{phdParrilo}). Many positive polynomials are not sums of squares of
polynomials following \cite{Blek}. However, some variants have been designed to make this
approach more general ; see e.g. the survey by~\cite{laurent2009sums} and references
therein. In a nutshell, the core and initial idea is as follows.

A polynomial $f$ is non-negative over $\R^n$ if it can be written as
an SOS $s_1^2+\cdots +s_r^2$ with $s_i \in \R[X]$ for
$1\leq i \leq r$. Also $f$ is non-negative over the semi-algebraic set
$S$ if it can be written as
$s_1^2+\cdots+s_r^2+\sum_{j=1}^m \sigma_j g_j$ where $\sigma_i$ is a
sum of squares in $\R[X]$ for $1\leq j \leq m$. It turns out that,
thanks to the ``Gram matrix method'' (see
e.g.~\cite{Choi95,Las01sos,phdParrilo}), computing such decompositions can be
reduced to solving Linear Matrix Inequalities (LMI). This boils down
to considering a semidefinite programming (SDP) problem.

For instance, on input $f \in \Q[X]$ of even degree $d = 2k$, the
decomposition $f=s_1^2+\cdots+s_r^2$ is a by-product of a
decomposition of the form $f = v_k^T L^T D L v_k$, where $v_k$ is the
vector of all monomials of degree $\leq k$ in $\Q[X]$, $L$ is a lower
triangular matrix with non-negative real entries on the diagonal and
$D$ is a diagonal matrix with non-negative real entries. The matrices
$L$ and $D$ are obtained after computing a symmetric matrix $G$ (the
Gram matrix), semidefinite positive, such that $f = v_k^T G v_k$. Such
a matrix $G$ is found using solvers for LMIs. Such inequalities can be
solved symbolically (see~\cite{Simone16}), but the degrees of the
algebraic extensions needed to encode exactly the solutions are
prohibitive on large examples \cite{NRS10}.
Besides, there exist fast numerical solvers for solving LMIs
implemented in double precision,
e.g.~SeDuMi by~\cite{Sturm98usingsedumi},~SDPA by~\cite{Yamashita10SDPA} as
well as arbitrary-precision solvers, e.g.~SDPA-GMP by~\cite{Nakata10GMP},
successfully applied in many contexts, including bounds for kissing
numbers by~\cite{Bachoc06newupper} or computation of (real) radical
ideals by~\cite{Las13ideal}.
%Among successfull applications of high-precision SDP solvers, we
%mention the use of SDPA-QD (implemented with quadruple precision
%floating point numbers), from the SDPA
%family~\cite{Yamashita10SDPA} in the context of
%computing upper bounds for the density of binary sphere packing.

But using solely numerical solvers yields ``approximate''
non-negativity certificates. In our example, the matrices $L$ and $D$
(and consequently the polynomials $s_1, \ldots, s_r$) are not known
exactly.

This raises topical questions. The first one is how to use symbolic computation
jointly with these numerical solvers to get {\it exact} certificates? Since not
all positive polynomials are SOS, what to do when SOS certificates do not exist?
Also, given inputs with rational coefficients, can we obtain certificates with
rational coefficients?

%In the univariate case, we can rely on semidefinite programming (SDP)
%instead of root isolation as in Algorithm~$\univsostwo$ designed
%in~\cite{Chevillard11} and analyzed in~\cite{univsos}.  
%
%For these questions, we inherit from previous contributions in the univariate case~\cite{Chevillard11,univsos} as well as in the multivariate case by Parillo and Peyrl~\cite{PaPe08} and next Kaltofen, Li, Yang and
%Zhi~\cite{KLYZ08}. 

For these questions, we inherit from contributions in the univariate
case by~\cite{Chevillard11,univsos} as well as in the multivariate
case by~\cite{PaPe08,KLYZ08}. Note that \cite{KLYZ08, Kaltofen12} allow us to
compute SOS with rational coefficients on some degenerate examples. Moreover,
\cite{Kaltofen12} allows to compute decompositions into sums of squares of
rational fractions. Diophantine aspects
are considered in \cite{Safey10Siam, GSZ13}. When an SOS decomposition exists
with coefficients in a totally real Galois field,~\cite{Hillar09} and \cite{Quarez2010}
provide bounds on the total number of squares.

In the univariate (un)-constrained case, given $f \in \Q[X]$, the algorithm
by~\cite{Chevillard11} computes an exact (weighted) SOS decomposition $f =
\sum_{i=1}^t c_i g_i^2$ with $c_i\in \Q$ and $g_i \in \Q[X]$. 
{We call such SOS
decompositions {\em weighted} because the coefficients $c_i$ are considered
outside the square, which helps when one wants to output data with rational
coefficients only}. To do that, the algorithm considers first a perturbation of
$f$, performs (complex) root isolation to get an approximate SOS decomposition
of $f$. When the isolation is precise enough, the algorithm relies the
perturbation terms to recover an exact rational decomposition.

In the multivariate unconstrained case, Parillo and Peyrl designed a
rounding-projection algorithm in~\cite{PaPe08} to compute a weighted rational
SOS decompositon of a given polynomial $f$ in the interior of the SOS cone. The
algorithm computes an approximate Gram matrix of $f$, and rounds it to a
rational matrix. With sufficient precision digits, the algorithm performs an
orthogonal projection to recover an exact Gram matrix of $f$. The SOS
decomposition is then obtained with an exact $L D L^T$ procedure. This approach
was significantly extended in~\cite{KLYZ08} to handle rational functions and in~\cite{Guo12} to derive certificates of impossibility for Hilbert-Artin representations of a given degree. 
{In a recent work {by}~\cite{Laplagne18}, the author derives an algorithm based on facial reduction techniques to obtain exact rational decompositions for some sub-classes of non-negative polynomials lying in the border of the SOS cone. 
Among such degenerate sub-classes, he considers polynomials that can be written as sums of squares of {polynomials} with coefficients in an algebraic extension of $\Q$ of odd degree.}

% project the result and succeeds  $L D L^T$ this matrix with 
%can perform  on an orthogonal projection of, allowing to recover an %exact rational decomposition. 

\textit{Main contributions.} This work provides an algorithmic framework for
computing %to handle (un)-constrained polynomial problems
exact rational weighted SOS decompositions in some favourable situations. The
first contribution, given in Section~\ref{sec:intsos}, is a hybrid
numeric-symbolic algorithm, called $\intsos$, providing rational SOS
decompositions for polynomials lying in the interior of the SOS cone. As for the
algorithm by~\cite{Chevillard11}, the main idea is to perturb the input
polynomial, then to obtain an approximate SOS decomposition (through some Gram
matrix of the perturbation by solving an SDP problem), and to recover an exact
decomposition using the perturbation terms.

In Section~\ref{sec:reznick}, we rely on $\intsos$ to compute decompositions of positive definite forms into SOS of rational functions, based on
Reznick's representations, yielding an algorithm, called
{$\reznicksos$}. {In Section~\ref{sec:hilbert}, we provide another algorithm, called $\hilbertsos$,  to decompose non-negative polynomials into SOS of rational functions, under the assumption that the numerator belongs to the interior of the SOS cone. 
}
{
In Section~\ref{sec:putinar}, we rely on $\intsos$ to
compute weighted SOS decompositions for polynomials positive over
basic compact semi-algebraic sets, yielding the $\putinarsos$ algorithm.\\
When the input is an $n$-variate polynomial of degree $d$ with integer
coefficients of maximum bit size $\tau$, we prove  in Section~\ref{sec:intsos} that
Algorithm~$\intsos$ runs in boolean time \revise{ $\tau^2  d^{d^{\bigo{ (n)  }}}$.
In the former estimate, the exponent $d^n$ can be replaced by the cardinality of the Newton polytope of the input polynomial}.
%outputs SOS polynomials of bit size bounded by $\tau  d^{\bigo{(n)}}$.
This also yields bit complexity analysis for Algorithm~$\reznicksos$ (see Section~\ref{sec:reznick}) and
Algorithm~$\putinarsos$ (see Section~\ref{sec:putinar}). To the best of our knowledge, these are the
first complexity estimates for the output of algorithms providing
exact multivariate SOS decompositions. 
%{The constants in the exponents are explicitely given in the sequel.}
\\
The three algorithms are implemented within a Maple procedure, called
$\multivsos$, integrated in the $\realcertify$  Maple library by~\cite{RealCertify}. In Section~\ref{sec:benchs}, we provide benchmarks to evaluate the
performance of $\multivsos$. We compare it with previous approaches
in~\cite{PaPe08} as well as with the more general methods based on the critical
point method and Cylindrical Algebraic Decomposition. \\
This paper is the follow-up of our previous contribution~\citep{multivsos18},
published in the proceedings of ISSAC'18. 
The main theoretical and practical novelties are the following: we provide explicit bounds for the bit complexity analyzes of our algorithms. 
In Section~\ref{sec:pp}, we state formally the rounding-projection algorithm from~\cite{PaPe08}, analyze its bit complexity and compare it with our algorithm~$\intsos$. We show that both algorithms have the same bit complexity.
Another novelty is in Section~\ref{sec:hilbert}, where we explain how to handle the sub-class of non-negative polynomials  admitting an Hilbert-Artin's representation, for which the numerator belongs to the interior of the SOS cone.
In Section~\ref{sec:ppPutinar}, we state a constrained version of the rounding-projection algorithm. Again, this algorithm has the same bit complexity as~$\putinarsos$. 
We have updated accordingly Section~\ref{sec:benchs} by providing some related numerical comparisons. 
We also consider benchmarks involving non-negative polynomials which do not belong to the interior of the SOS cone.
}

\paragraph*{Acknowledgments.} M.~Safey El Din is supported by the
ANR-18-CE33-0011 {\sc Sesame}, the
ANR-19-CE40-0018 {\sc De Rerum Natura} projects funded by ANR (Agence Nationale
de la Recherche), the ANR-19-CE48-0015 {\sc Ecarp} project funded jointly by
ANR and FWF, and the {\sc CAMiSAdo} project funded by
PGMO/FMJH. V.~Magron benefited from the support of the FMJH Program PGMO (EPICS project) and from the support of EDF, Thales, Orange et Criteo, as well as from the Tremplin ERC Stg Grant ANR-18-ERC2-0004-01 (T-COPS project).
Both authors are supported by European Union's Horizon 2020 research and innovation programme under the Marie Sklodowska-Curie grant agreement 813211 (POEMA).
The authors would like to thank David Papp and Maria Macaulay for their useful comments who led to the present version, where complexity estimates have been fixed. 
%
%\textit{Related Works.}
%computation of (real) radical ideals. 
%Preliminary experiments emphasize competitive results when the bit size of the polynomials is small, e.g.~for power sums of degree up to 1000. However, the performance decrease when the bit size becomes
%larger, either for polynomial benchmarks from~\cite{Chevillard11} or
%modified Wilkinson polynomials. 

%\textit{Plan.} This paper is structured as follows. In Section~\ref{sec:prelim}, we 
%

%%% Local Variables:
%%% mode: latex
%%% TeX-master: "multivsos"
%%% End:

%!TEX root = multivsosJSC.tex
\section{Preliminaries}
\label{sec:prelim}

Let $\Z$ be the ring of integers and $X = (X_1, \ldots, X_n)$. For $\alpha =
(\alpha_1,\dots,\alpha_n) \in \N^n$, one has $|\alpha| := \alpha_1 + \dots +
\alpha_n$ and $X^\alpha := X_1^{\alpha_1} \dots X_n^{\alpha_n}$. For all $k \in
\N$, we let $\N^{n}_k := \{ \alpha \in \N^{n} : |\alpha| \leq k \}$, whose
cardinality is the binomial $\binom{n+k}{k}$. A polynomial $f \in \R[X]$ of
degree $d = 2k$ is written as $ f \,=\,\sum_{|\alpha|\leq d} \, f_{\alpha} \,
X^\alpha $ and we identify $f$ with its vector of coefficients $\f=(f_{\alpha})$
in the basis $(X^\alpha)$, $\alpha \in\N_d^n$. {When referring to univariate
  polynomials, we use the indeterminate $E$ and we denote by $\Z[E]$ the set of
  univariate polynomials with integer coefficients.} Let ${\Sigma}[X]$ be the
convex cone of sums of squares in $\R[X]$ and $\mathring{\Sigma}[X]$ be the
interior of $\Sigma[X]$. We will be interested in those polynomials which lie in
$\Z[X] \cap \Sigma[X]$. 
% We note $\Sigma_\Z(X) := \Z[X] \cap \Sigma[X]$ and
% $\mathring{\Sigma}_\Z[X] := \Z[X] \cap \mathring{\Sigma}[X]$ . 
For instance, the
polynomial
\[f = 4 X_1^4 + 4
X_1^3 X_2 - 7 X_1^2 X_2^2 - 2 X_1 X_2^3 + 10 X_2^4 = (2 X_1 X_2 + X_2^2)^2 + (2
X_1^2 + X_1 X_2 - 3 X_2^2)^2\] lies in $\Z[X] \cap \Sigma[X]$. 
%Similarly, we define $\Sigma_\Q(X)$ and $\mathring{\Sigma}_\Q[X]$.
%For a subring
%$\mathbb{K}$ of $\R$, we denote by $\Sigma_{\mathbb{K}}(X)$ the %subset
%of polynomials in $\Sigma[X]$ with coefficients in $\mathbb{K}$ and %by
%$\mathring{\Sigma}_{\mathbb{K}}$ its interior. 

%\textit{Notation for complexity estimates.} 
The complexity estimates in this paper rely on the bit complexity model. The bit
size of an integer $b$ is denoted by $\tau(b) := \lfloor \log_2 (|b|) \rfloor +
1$ with $\tau(0) := 1$. For $f = \sum_{|\alpha| \leq d} f_\alpha X^\alpha \in
\Z[X]$ of degree $d$, we denote $\|f\|_\infty := \max_{|\alpha| \leq d}
|f_\alpha|$ and $\tau(f) := \tau(\|f\|_\infty)$ with slight abuse of notation.
Given $b \in \Z$ and $c \in \Z \backslash \{0\}$ with gcd$(b,c) = 1$, we define
$\tau(b/c) := \max \{\tau(b), \tau(c)\}$. For two mappings $g,h : \N^l \to \R$,
we use the notation ``$g(v) = \bigo{(h(v))}$'' to state the existence of $b \in
\N$ such that $g(v) \leq b h(v)$, for all $v \in \N^l$.

The {\em Newton polytope} or {\em cage} $\polytope{(f)}$ is the convex hull of
the vectors of exponents of monomials that occur in $f \in \R[X]$. For the above
example, $\polytope{(f)} = \{(4,0),(3,1),(2,2),(1,3),(0,4)\}$. For a symmetric
real matrix $G$, we note $G \succeq 0$ (resp.~$G \succ 0$) when $G$ has only
non-negative (resp.~positive) eigenvalues and we say that $G$ is {\em positive
  semidefinite} (SDP) (resp.~{\em positive definite}).
The minimal eigenvalue of a real symmetric matrix $G$ is denoted by $\lambda_{\min}(G)$.
{For a given Newton polytope $P$, let $\Sigma_P[X]$ be the convex cone of sums of
squares whose Newton polytope is contained in $P$. 
Since the Newton polytope $P$ is often clear from the context, we suppress the index $P$.}

\if{
Given a real sequence
$\y =(y_{\alpha})_{\alpha \in \N^n}$, the linear functional
$\ell_\y : \R[X] \to \R$ is defined by
$\ell_\y(f) := \sum_{|\alpha| \leq d} f_{\alpha} y_{\alpha}$, for all
$f \in \R[X]$. 
We associate to $\y$ the {\it moment matrix} $M_k(\y)$, that is the real
symmetric matrix with rows (resp.~columns) indexed by $\N_k^n$ and the
following entrywise definition:
$(M_k(\y))_{\beta,\gamma} := \ell_\y(X^{\beta + \gamma})$ forall
$\beta, \gamma \in \N_k^n$.

Given $g \in \R[X]$, we associate to $\y$ the {\it localizing matrix},
that is the real symmetric matrix $M_k(g \, \y)$ with rows (resp.~
columns) indexed by $\N_k^{n}$ and the following entrywise definition:
$(M_k(g \, \y))_{\beta, \gamma} := \ell_\y(g \, X^{\beta + \gamma}) $
forall $\beta, \gamma \in \N_k^n $.
\begin{equation}
\label{eq:primalsdp}
\begin{aligned}
\inf\limits_{\y} \quad & \sum_{|\alpha| \leq d} f_\alpha y_{\alpha}   \quad 
\text{s.t.} \quad & M_k(\y) \succeq 0 \,, \quad y_0 = 1  \,. \\
\end{aligned}
\end{equation}
Writing $M_k(\y) = \sum_{|\alpha| \leq d} y_\alpha B_\alpha$, the dual of SDP~\eqref{eq:primalsdp} is also an SDP:
}\fi

\revise{
With $f \in \R[X]$ of degree $d = 2k$, we consider the SDP feasibility program:
\begin{equation}
\label{eq:dualsdp}
\begin{aligned}
\text{Find } G \succeq 0  & \quad 
\text{s.t.} \quad & \trace{(G \, B_\gamma)} = f_\gamma \,, \quad \forall \gamma \in \N_d^n \,, \\
\end{aligned}
\end{equation}
where $\trace(M)$ (for a given matrix $M$) denotes the trace of $M$, $B_\gamma$  has rows (resp.~
columns) indexed by $\N_k^{n}$ with $(\alpha, \beta)$ entry equal to 1 if $\alpha + \beta = \gamma$ and 0 otherwise.
When $f \in \Sigma[X]$, SDP \eqref{eq:dualsdp} has a feasible solution $G = \sum_{i=1}^r \lambda_i \q_i \, \q_i^T $, with the $\q_i$
being the eigenvectors of $G$ corresponding to the non-negative
eigenvalues $\lambda_i$, for all $i=1,\dots,r$, and 
$f  = \sum_{i=1}^r \lambda_i q_i^2$. 
For more details, see, e.g., \cite{phdParrilo,Las01sos}.}
%When $f \in \Sigma[X]$, 
%The next statement follows directly from~.
\if{
\begin{theorem}\cite[Theorem~3.2]{Las01sos}
\label{th:lasunconstrained}
Let $f \in \R[X]$ of degree $d = 2k$ and global infimum
$f^\star := \inf_{\bmx \in \R^n} f(\bmx)$ 
\revise{ and assume that $f - f^\star \in \Sigma[X]$. 
%then the optimal value of SDP~\eqref{eq:primalsdp} is equal to $f^\star$.  Conversely, 
Then SDP~\eqref{eq:dualsdp} has a feasible solution
$G^\star = \sum_{i=1}^r \lambda_i \q_i \, \q_i^T $, with the $\q_i$
being the eigenvectors of $G^\star$ corresponding to the non-negative
eigenvalues $\lambda_i$, for all $i=1,\dots,r$, and 
$f - f^\star = \sum_{i=1}^r \lambda_i q_i^2$.}
\end{theorem}
}\fi
For the sake of efficiency, one reduces the size of the matrix $G$
by indexing its rows and columns by half of $\polytope{(f)}$:
\begin{theorem}\cite[Theorem~1]{Reznick78}
\label{th:np}
Let $f \in \Sigma[X]$ with $f = \sum_{i=1}^r s_i^2$, $P := \polytope{(f)}$ and $Q := P/2 \cap \N^n$.  Then for all $i=1,\dots,r$, $\polytope{(s_i)} \subseteq Q$.
\end{theorem}
\revise{
Then from now on, we consider SDP \eqref{eq:dualsdp} with $G$ (resp.~each matrix $B_\gamma$) having its rows and columns indexed by $Q$. 
}

Given $f \in \R[X]$, one can theoretically certify that $f$ lies in
$\Sigma[X]$ by solving SDP~\eqref{eq:dualsdp}.  However,
available SDP solvers are typically implemented in finite-precision
and require the existence of a strictly feasible solution $G \succ 0$
to converge. This is equivalent for $f$ to lie in
$\mathring{\Sigma}[X]$ as stated in~\cite[Proposition~5.5]{Choi95}:
\begin{theorem}
\label{th:intsospdGram}
Let $f \in \Z[X]$ with $P := \polytope{(f)}$, $Q := P/2 \cap \N^n$ and $v_k$ be the
vector of all monomials with support in $Q$. Then $f \in \mathring{\Sigma}[X]$
if and only if there exists a positive definite matrix $G$ such that
$f = v_k^T G v_k$.
\end{theorem}
Eventually, we will rely on the following bound for the roots of polynomials with integer coefficients:
\begin{lemma}{~\cite[Theorem~4.2 (ii)]{Mignotte1992}}
\label{th:cauchyint}
Let $f \in \Z[E]$ of degree $d$, with coefficient bit size bounded from above by $\tau$. If $f(e) = 0$ and $e \neq 0$, then $\frac{1}{2^\tau + 1} \leq |e| \leq 2^\tau + 1$.
\end{lemma}
%
%%% Local Variables:
%%% mode: latex
%%% TeX-master: "multivsos"
%%% End:
%Hence, we focus now on the case of polynomials lying in
%$\mathring{\Sigma}[X]$.
%!TEX root = multivsosJSC.tex
\section{Exact SOS representations}
\label{sec:intsos}
The aim of this section is to state and analyze a hybrid numeric-symbolic
algorithm, called $\intsos$, computing weighted SOS decompositions of
polynomials in $\Z[X]\cap \mathring{\Sigma}[X]$. This algorithm relies on perturbations of such polynomials. {We first establish the following preliminary results.}
\revise{
\begin{proposition}
\label{th:boundGram}
Let $f \in \Z[X]\cap \mathring{\Sigma}[X]$ of degree $d = 2k$, with
$\tau = \tau(f)$, $P=\polytope(f)$ and $Q := P/2 \cap \N^n$. 
Then, there exists a feasible $G \succ 0$ for SDP \eqref{eq:dualsdp} and positive integers $R, N$ such that $\sqrt{\trace{(G^2)}} \leq R$ and $\lambda_{\min} (G) > 2^{-N}$, with  $\tau(R), N \leq \tau  d^{d^{\bigo{ (n)  }}}$.
\end{proposition}

\begin{proof}
Let $v_k$ be the vector of all monomials $X^\alpha$, with $\alpha$ in $Q$. Note
  that each monomial in $v_k$ has degree $\leq k$ and that
  $v_k^T v_k = \sum_{\alpha \in Q} X^{2 \alpha}$. Since
  $f \in \mathring{\Sigma}[X]$, there exists by Theorem~\ref{th:intsospdGram} a 
  matrix $G \succ 0$ such that $f =v_k^T G v_k$, with positive smallest eigenvalue.
%The bound on $\sqrt{\trace{(G^2)}}$ follows directly from  .
Let us note $\nsdp$ the size of $G$ and $\msdp$ the number of equality constraints of SDP \eqref{eq:dualsdp}.
Then \cite[Theorem 3.1 (i)]{porkolab1997complexity} implies that $\sqrt{\trace{(G^2)}} \leq R$, with $\tau(R) = \tau \nsdp^{\bigo{(\msdp)} }$.
Using that for all $k \geq 2$,
\[
\nsdp \leq \binom{n+k}{n} = \frac{(n+k) \cdots (k+1)}{n!} =
(1+\frac{k}{n})(1+\frac{k}{n-1})\cdots(1+k)     \leq k^{n-1} (1+k) \leq 2 k^n \leq d^n \,,
\]
and $\msdp \leq \binom{n+d}{n} \leq 2 d^n$, we obtain the desired bit estimate for $R$, as well for entries of $G$. 
Then note that the minimal eigenvalue is a root of the univariate polynomial $\det (G - E I) \in \Q[E]$. 
The coefficients of this latter polynomial are obtained by $\nsdp !$ products of numbers of bit size upper bounded by $\tau  d^{d^{\bigo{ (n)  }}}$, thus their bit size is also upper bounded by $\tau  d^{d^{\bigo{ (n)  }}}$. 
Then Lemma \ref{th:cauchyint} yields the desired result.
%a product The bit size of this determinant is upper bounded by $\nsdp$
%Similarly, by using again \cite{porkolab1997complexity}, 
%the smallest eigenvalue of $G$ has bit size upper bounded by  $\tau  d^{\bigo{(d^n)}}$.
\end{proof}

}
\begin{proposition}
\label{th:boundeps}
Let $f$ as above. 
Then, there exists
$N\in \N-\{0\}$ such that for $\varepsilon := \frac{1}{2^N}$, 
$f - \varepsilon \sum_{\alpha \in Q} X^{2 \alpha}
\in \mathring{\Sigma}[X]$, with %{$N \leq \tau(\varepsilon) \leq \bigo{( \tau \cdot (4 d+2)^{3n+3} )}$}.
\revise{ $N \leq \tau(\varepsilon) \leq  d^{d^{\bigo{ (n)  }}}$.}
\end{proposition}

\if{
The proof of this result relies on the following technical statement whose proof
is postponed in the Appendix. 

\begin{proposition}
  \label{prop:ori-cpm}
  Let $g\in \Z[X]$ of degree $d$ and let $\tau = \tau(g)$. Assume that the
  algebraic set $V(g)\subset \mathbb{C}^n$ defined by $g = 0$ is smooth. Then, there
  exists a polynomial $w\in \Z[X_1]$ of degree $\leq d^n$ with coefficients of
  bit size $\leq \tau \cdot (4 d+2)^{3n}$ such that its set of real roots
  contains the critical values of the restriction of the projection on the
  $X_1$-axis to $V(g)$.
\end{proposition}

\begin{proof}[Proof of Proposition~\ref{th:boundeps}]
}\fi

\begin{proof}
As in the proof of Proposition \ref{th:boundGram}, since
  $f \in \mathring{\Sigma}[X]$, there exists by Theorem~\ref{th:intsospdGram} a 
  matrix $G \succ 0$ such that $f =v_k^T G v_k$, with positive smallest eigenvalue
  $\lambda$. Let us define
  $N := \lceil \log_2 \frac{1}{\lambda} \rceil + 1$, i.e.~the smallest
  integer such that
  $\varepsilon = \frac{1}{2^N} \leq \frac{\lambda}{2}$. Then,
  $\lambda > \varepsilon$ and the matrix $G - \varepsilon I$ has only
  positive eigenvalues. 
  Hence, one has
\[
f_\varepsilon := f - \varepsilon \sum_{\alpha \in Q} X^{2 \alpha}
  = v_k^T G v_k - \varepsilon v_k^T I v_k = v_k^T (G - \varepsilon I)
  v_k \,,\]
  yielding $f_\varepsilon \in \mathring{\Sigma}[X]$. 
The upper bound on the bit size on $N$ follows directly from Proposition \ref{th:boundGram}.
\end{proof}

\if{
\begin{proof}
  Let $v_k$ be the vector of all monomials $X^\alpha$, with $\alpha$ in $Q$. Note
  that each monomial in $v_k$ has degree $\leq k$ and that
  $v_k^T v_k = \sum_{\alpha \in Q} X^{2 \alpha}$. Since
  $f \in \mathring{\Sigma}[X]$, there exists by Theorem~\ref{th:intsospdGram} a 
  matrix $G \succ 0$ such that $f =v_k^T G v_k$, with positive smallest eigenvalue
  $\lambda$. Let us define
  $N := \lceil \log_2 \frac{1}{\lambda} \rceil + 1$, i.e.~the smallest
  integer such that
  $\varepsilon = \frac{1}{2^N} \leq \frac{\lambda}{2}$. Then,
  $\lambda > \varepsilon$ and the matrix $G - \varepsilon I$ has only
  positive eigenvalues. 
  Hence, one has
\[
f_\varepsilon := f - \varepsilon \sum_{\alpha \in Q} X^{2 \alpha}
  = v_k^T G v_k - \varepsilon v_k^T I v_k = v_k^T (G - \varepsilon I)
  v_k \,,\]
  yielding $f_\varepsilon \in \mathring{\Sigma}[X]$.  

\if{  
  For the second
  claim, 
\if{  
  it is enough to select $\varepsilon := \frac{1}{2^N}$ such that $\varepsilon \leq \lambda = \min_{\|v \|_2 = 1} \sqrt{v^T G G v}$. Moreover, one has $f = v_k^T G v_k = \langle G v_k, v_k \rangle \leq  \sqrt{v_k^T G G v_k} \|v_k\|_2$, thanks to the Cauchy-Schwartz inequality. Thus, it is enough to select $\varepsilon \leq \min \{ f(\bmx) : \|v_k(\bmx)\|_2 = 1 \}$.}\fi let us consider the algebraic set $V$ defined by
  \[
  f(X) - E = \frac{\partial f}{\partial X_1} = \dots = \frac{\partial f}{\partial X_n}  \,.
  \]
Let us note $A$ the projection of $V \cap \R^n$ on the $E$-axis.
Note that $A$ contains the minimizers of $f(\bmx)$. 
%on the set $\{x\in\R^n : v_k^T (\bmx) v_k(\bmx) = 1 \}$.

{
Note that the algebraic set defined by $f - E = 0$ is smooth (since it is a graph).
Applying Proposition~\ref{prop:ori-cpm} to $f-E$}, we deduce that there exists a polynomial in $\Z[E]$ of degree less than $d^{n+1}$ with coefficients of bit size less than $\tau \cdot (4 d + 2)^{3n+3}$ such that its set of real roots contains $A$. 
}\fi

By {Lemma}~\ref{th:cauchyint}, it follows that it is enough to select $N \leq \bigo{( \tau \cdot (4 d+2)^{3n+3} )}$.

%  Let us consider the set
%  \[A := \{e \in \R^{> 0} : \forall \bmx \in \R^n, f(\bmx) - e \sum_{\alpha
%    \in Q} \bmx^{2 \alpha} \geq 0\} \,. \] 
    %

%Using \cite[Thm 14.16]{BPR06}, $A$ is defined by univariate polynomials of degree in $d^{\bigo{(n)}}$ with coefficients of bit size bounded by $\tau d^{\bigo{(n)}}$. Hence the bit size of the mimimum absolute value of their non-zero real roots is below bounded by $\tau d^{\bigo{(n)}}$.
\end{proof}
}\fi
The following can be found in~\cite[Lemma~2.1]{Bai89}
and~\cite[Theorem~3.2]{Bai89}.
\begin{proposition}
\label{th:boundchol}
Let ${G} \succ 0$ be a matrix with
rational entries indexed on $\N^n_r$. Let $L$ be the factor of ${G}$
computed using Cholesky's decomposition with finite precision
$\delta_c$. Then $L L^T = {G} + F$ where
\begin{align}
\label{eq:chol1}
|F_{\alpha, \beta}| \leq \frac{ {(r+1)2^{-\delta_c}} |{G}_{\alpha,\alpha} \, {G}_{\beta,\beta}|^{\frac{1}{2}}}{1 - (r+1)2^{-\delta_c}}  \,.
\end{align}
In addition, if the smallest eigenvalue 
$\tilde{\lambda}$ of ${G}$ satisfies the  inequality
\begin{align}
\label{eq:chol2}
2^{-\delta_c} < \frac{ \tilde{\lambda} }{ r^2 + r + (r-1) \tilde{\lambda}} \,,
\end{align}
Cholesky's decomposition returns a rational nonsingular factor $L$.
\end{proposition}

\subsection{Algorithm~\texttt{intsos}}
\label{sec:algo}
We present our algorithm $\intsos$ computing exact weighted rational SOS
decompositions for polynomials in $\Z[X]\cap \mathring{\Sigma}[X]$. 

{
\begin{algorithm}
\caption{$\intsos$%: algorithm to compute weighted SOS decompositions of multivariate polynomials in $\mathring{\Sigma}[X]$.
}
\label{alg:intsos}
\begin{algorithmic}[1]
%\Require $f \in \mathring{\Sigma}_\Z[X]$, positive $\varepsilon \in \Q$, precision parameters $\delta, R \in \N$ for the SDP solver 
\Require $f \in \Z[X]$, positive $\varepsilon \in \Q$, precision parameters $\delta, R \in \N$ for the SDP solver, precision $\delta_c \in \N$ for the Cholesky's decomposition
\Ensure list $\clist$ of numbers in $\Q$ and list $\slist$ of polynomials in $\Q[X]$
\State $P := \polytope{(f)}$, $Q := P/2 \cap \N^n$ \label{line:np}
\State $t := \sum_{\alpha \in Q} X^{2 \alpha}$, $f_\varepsilon \gets f - \varepsilon t$
\While {$f_\varepsilon \notin \mathring{\Sigma}[X]$} \label{line:epsi}
 $\varepsilon \gets \frac{\varepsilon}{2}$, $f_\varepsilon \gets f - \varepsilon t$
\EndWhile \label{line:epsf}
\State ok := false
\While {not ok} \label{line:deltai}
\State $(\tilde{G}, \tilde{\lambda}) \gets \sdpfun{f_\varepsilon}{ \delta}{R}$ \label{line:sdp}
\State $(s_1,\dots,s_r) \gets \choleskyfun{\tilde{G}}{\tilde{\lambda}}{\delta_c}$ \label{line:chol} \Comment{$f_\varepsilon \simeq \sum_{i=1}^r  s_i^2$}
\State $u \gets f_\varepsilon - \sum_{i=1}^r  s_i^2$
\State $\clist \gets [1,\dots,1] $, $\slist \gets [s_1,\dots,s_r]$
\For {$\alpha \in Q$}  $\varepsilon_{\alpha} := \varepsilon$
\EndFor
\State $\clist, \slist,(\varepsilon_\alpha) \gets \absorbfun{u}{Q}{(\varepsilon_\alpha)}{\clist}{\slist}$ \label{line:absorb}
\If {$\min_{\alpha \in Q} \{ \varepsilon_\alpha \} \geq 0$} ok := true \label{line:sosok}
\Else $\ \delta \gets 2 \delta$, $R \gets 2 R$, $\delta_c \gets 2 \delta_c$
\EndIf
\EndWhile \label{line:deltaf}
\For {$\alpha \in Q$}   $\clist \gets  \clist \cup \{ \varepsilon_\alpha \}$, $\slist \gets  \slist \cup \{ X^\alpha \}$
\EndFor
\State \Return $\clist$, $\slist$
\end{algorithmic}
\end{algorithm}
}

{
\begin{algorithm}
\caption{$\absorb$%: algorithm to compute weighted SOS decompositions of multivariate polynomials in $\mathring{\Sigma}[X]$.
}
\label{alg:absorb}
\begin{algorithmic}[1]
\Require $u \in \Q[X]$, multi-index set $Q$, lists $(\varepsilon_\alpha)$ and $\clist$ of numbers in $\Q$, list $\slist$ of polynomials in $\Q[X]$
\Ensure lists $(\varepsilon_\alpha)$ and $\clist$ of numbers in $\Q$, list $\slist$ of polynomials in $\Q[X]$
\For {$\gamma \in \spt{u}$} \label{line:absi}
\If {$\gamma \in (2\N)^n$} $\alpha := \frac{\gamma}{2}$, $\varepsilon_\alpha := \varepsilon_\alpha + u_\gamma$ \label{line:intsoseven}
\Else \State Find $\alpha$, $\beta \in Q$ such that $\gamma = \alpha + \beta $ \label{line:intsosodd}
\State $\varepsilon_\alpha := \varepsilon_\alpha - \frac{|u_\gamma|}{2}$, $\varepsilon_\beta := \varepsilon_\beta - \frac{|u_\gamma|}{2}$
\State $\clist \gets  \clist \cup \{ \frac{|u_\gamma|}{2} \}$
\State $\slist \gets  \slist \cup \{ X^\alpha + \sgn{(u_\gamma)} X^\beta \}$
\EndIf
\EndFor \label{line:absf}
\end{algorithmic}
\end{algorithm}
}
%Given $f \in \mathring{\Sigma}_\Z[X]$ of degree $d = 2k$, one first 

Given $f \in \Z[X]$ of degree $d = 2k$, one first computes its Newton polytope
$P := \polytope{(f)}$ (see line~\lineref{line:np}) and $Q := P/2 \cap \N^n$
using standard algorithms such as quickhull by~\cite{Barber96}. The loop going from
line~\lineref{line:epsi} to line~\lineref{line:epsf} finds a positive
$\varepsilon \in \Q$ such that the perturbed polynomial $f_\varepsilon := f -
\varepsilon \sum_{\alpha \in Q} X^{2 \alpha}$ is also in $\mathring{\Sigma}[X]$.
This is done thanks to any external {\em oracle} deciding the non-negativity of a polynomial. 
{Even if this oracle is able to {\em decide} non-negativity, we would like to emphasize that our algorithm outputs an SOS certificate in order to {\em certify} the non-negativity of the input. 
In practice, we often choose the value of $\varepsilon$ while relying on a heuristic technique rather than this external oracle, for the sake of efficiency (see Section~\ref{sec:benchs} for more details).}

\if{
If $f \in \Z[X]\cap \mathring{\Sigma}[X]$, then the set $\{e \in \R^{>
  0} : \forall \bmx \in \R^n, f(\bmx) - e \sum_{\alpha \in Q} \bmx^{2 \alpha}
\geq 0\}$ is non empty (see the proof of Proposition~\ref{th:boundeps}). 
If the oracle asserts that $\bmx \mapsto f(\bmx) - e \sum_{\alpha \in Q} \bmx^{2 \alpha}$ is non-negative on $\R^n$, then $e$ belongs to this set and it is enough to select $\varepsilon = e / 2$ to ensure that $f_\varepsilon := f - \varepsilon \sum_{\alpha \in Q} X^{2 \alpha} \in \mathring{\Sigma}[X]$.
}\fi
\if{
If $f \in \Z[X]\cap \mathring{\Sigma}[X]$, the existence of $\varepsilon$
is ensured as in the proof of Proposition~\ref{th:boundeps} if the set $\{e \in \R^{>
  0} : \forall \bmx \in \R^n, f(\bmx) - e \sum_{\alpha \in Q} \bmx^{2 \alpha}
\geq 0\}$ is non empty.
}\fi

Next, we enter in the loop starting from line~\lineref{line:deltai}.  Given $f_\varepsilon \in \Z[X]$, positive integers $\delta$
%$f_\varepsilon \in \mathring{\Sigma}[X]$, positive integers $\delta$
and $R$, the $\sdp$ function calls an SDP solver and tries to
compute a rational approximation $\tilde{G}$ of the Gram matrix
associated to $f_\varepsilon$ together with a rational approximation
$\tilde{\lambda}$ of its smallest eigenvalue. 
%Note that this is enough to index $\tilde{G}$ on $Q$ by Theorem~\ref{th:np}.

In order to analyse the complexity of the procedure (see
Remark~\ref{rk:ellipsoidIP}), we assume that $\sdp$ relies on the ellipsoid
algorithm by~\cite{GroetschelLovaszSchrijver93}.
\begin{remark}
\label{rk:ellipsoidIP}
In~\cite{deKlerkSDP}, the authors analyze the complexity of the short step,
primal interior point method, used in SDP solvers. Within fixed accuracy, they
obtain a polynomial complexity, as for the ellipsoid method, but the exact value
of the exponents is not provided.

Also, in practice, we use an
arbitrary-precision SDP solver implemented with an interior-point method.
\end{remark}
SDP problems are solved with this latter algorithm in polynomial-time within a
given accuracy $\delta$ and a radius bound $R$ on the Frobenius norm of $\tilde{G}$.
%of the matrix variable. 
%For given integers $\delta$ and $R$, 
The first
step consists of solving SDP~\eqref{eq:dualsdp} by computing an
approximate Gram matrix $\tilde{G} \succeq 2^{- \delta} I $ such that
\[
|\trace{( \tilde{G} B_\gamma)} - (f_\varepsilon)_{\gamma}| = | \sum_{\alpha+\beta = \gamma} \tilde{G}_{\alpha,\beta} -
(f_\varepsilon)_{\gamma}| \leq
2^{-\delta}
\]
and $\sqrt{\trace{(\tilde{G}^2)}} \leq R$.
We pick large enough integers $\delta$ and $R$ to obtain $\tilde{G} \succ 0$ and $\tilde{\lambda} > 0$ when $f_\varepsilon \in \mathring{\Sigma}[X]$.
%Otherwise, the process is repeated
%  with $2 \delta$ and $2 R$.

%The $\cholesky$ function takes as input 
%$\tilde{G}$, $\tilde{\lambda}$ and $\delta_c$. 
The $\cholesky$ function computes the approximate
Cholesky's decomposition $L L^T$ of $\tilde{G}$ with precision
$\delta_c$. In order to guarantee that $L$ will be a rational
nonsingular matrix, a preliminary step consists of verifying that the
inequality~\eqref{eq:chol2} holds, which happens when $\delta_c$
is large enough. Otherwise, $\cholesky$  selects the
smallest $\delta_c$ such as~\eqref{eq:chol2} holds.  Let $v_k$ be the size $r$ 
vector of all monomials $X^\alpha$ with $\alpha$ belonging to $Q$. The output is a list of
rational polynomials $[s_1,\dots,s_r]$ such that for all
$i=1,\dots,r$, $s_i$ is the inner product of the $i$-th row of $L$
by $v_k$. 
%By Theorem~\ref{th:lasunconstrained}, 
One would expect to have $f_\varepsilon = \sum_{i=1}^r s_i^2$ with $s_i \in \R[X]$ after using exact SDP and Cholesky's decomposition. Here, we have to consider  the remainder $u = f - \varepsilon \sum_{\alpha \in Q} X^{2 \alpha} - \sum_{i=1}^r s_i^2$, with $s_i \in \Q[X]$. 

After these steps, the algorithm starts to
perform symbolic computation with the $\absorb$ subroutine at
line~\lineref{line:absorb}. The loop from $\absorb$ is designed to obtain an
exact weigthed SOS decomposition of $\varepsilon t + u = \varepsilon
\sum_{\alpha \in Q} X^{2 \alpha} + \sum_{\gamma} u_\gamma X^\gamma$, yielding in
turn an exact decomposition of $f$. Each term $u_\gamma X^\gamma$ can be written
either $u_\gamma X^{2 \alpha}$ or $u_\gamma X^{\alpha + \beta}$, for $\alpha,
\beta \in Q$. In the former case (line~\lineref{line:intsoseven}), one has
\[
\varepsilon X^{2 \alpha} + u_\gamma X^{2 \alpha} = (\varepsilon + u_\gamma) X^{2 \alpha}
\,.\]
%which happens to be SOS when $\varepsilon + u_\gamma \geq 0$. 
In the latter case (line~\lineref{line:intsosodd}), one has 
\[
\varepsilon (X^{2 \alpha} +  X^{2 \beta}) + u_\gamma X^{\alpha + \beta} = |u_{\gamma}|/2 (X^{\alpha} + \sgn{(u_\gamma)} X^{\beta})^2 + (\varepsilon - |u_{\gamma}|/2) (X^{2 \alpha} + X^{2 \beta}) \,.
\]
%, which happens to be SOS when $\varepsilon \geq |u_{\gamma}|/2$. 
If the positivity test of line~\lineref{line:sosok} fails, then the coefficients of $u$ are too large and one cannot ensure that $\varepsilon t + u$ is SOS. So we repeat the same procedure after increasing the precision of the SDP solver and Cholesky's decomposition.

In prior work~\cite{univsos}, the authors and Schweighofer formalized and analyzed an algorithm called $\univsostwo$, initially provided in~\cite{Chevillard11}. Given a univariate polynomial $f > 0$ of degree $d = 2k$, this algorithm computes weighted SOS decompositions of $f$. With $t := \sum_{i=0}^k X^{2 i}$, the first numeric step of $\univsostwo$ is to find $\varepsilon$ such that the perturbed polynomial $f_\varepsilon := f - \varepsilon t > 0$ and to compute its complex roots, yielding an approximate SOS decomposition $s_1^2 + s_2^2$. The second symbolic step is very similar to the loop from line~\lineref{line:absi} to line~\lineref{line:absf} in $\intsos$: one considers the remainder polynomial $u := f_\varepsilon - s_1^2 - s_2^2$ and tries to computes an exact SOS decomposition of $\varepsilon t + u$. This succeeds for large enough precision of the root isolation procedure.
Therefore, $\intsos$ can be seen as an extension of $\univsostwo$ in the multivariate case by replacing the numeric step of root isolation by SDP and keeping the same symbolic step.

\begin{example}
\label{ex:intsos}
We apply Algorithm~$\intsos$ on \[f = 4 X_1^4 + 4 X_1^3 X_2 - 7 X_1^2 X_2^2 - 2
  X_1 X_2^3 + 10 X_2^4,\] with $\varepsilon = 1$, $\delta = R = 60$ and
$\delta_c = 10$. Then \[Q := \polytope{(f)}/2 \cap \N^n = \{(2,0),(1,1),(0,2)\}\]
(line~\lineref{line:np}). The loop from line~\lineref{line:epsi} to
line~\lineref{line:epsf} ends and we get $f - \varepsilon t = f - (X_1^4 + X_1^2
X_2^2 + X_2^2) \in \mathring{\Sigma}[X]$. The $\sdp$ (line~\lineref{line:sdp})
and $\cholesky$ (line~\lineref{line:chol}) procedures yield \[s_1 = 2 X_1^2+ X_1
X_2- \frac{8}{3} X_2^2, \quad s_2 = \frac{4}{3} X_1 X_2+ \frac{3}{2} X_2^2\quad
\text{and}\quad s_3
= \frac{2}{7} X_2^2.\] The remainder polynomial is $u = f- \varepsilon t - s_1^2
- s_2^2 - s_3^2 = - X_1^4- \frac{1}{9} X_1^2 X_2^2- \frac{2}{3} X_1 X_2^3-
\frac{781}{1764} X_2^4$.

At the end of the loop from line~\lineref{line:absi} to
line~\lineref{line:absf}, we obtain $\varepsilon_{(2,0)} = (\varepsilon - X_1^4
= 0$, which is the coefficient of $X_1^4$ in $\varepsilon t + u$. Then,
\[\varepsilon (X_1^2 X_2^2 + X_2^4) - \frac{2}{3} X_1 X_2^3 = \frac{1}{3} (X_1
X_2 - X_2^2)^2 + (\varepsilon - \frac{1}{3}) (X_1^2 X_2^2 + X_2^4).\] In the
polynomial $\varepsilon t + u$, the coefficient of $X_1^2 X_2^2$ is
$\varepsilon_{(1,1)} = \varepsilon - \frac{1}{3} - \frac{1}{9} = \frac{5}{9}$
and the coefficient of $X_4^4$ is $\varepsilon_{(0,2)} = \varepsilon -
\frac{1}{3} - \frac{781}{1764} = \frac{395}{1764}$.

Eventually, we obtain the weighted rational SOS decomposition: 
\begin{align*}
4 X_1^4 + 4 X_1^3 X_2 - 7 X_1^2 X_2^2 - 2 X_1 X_2^3 + 10 X_2^4 = &
  \frac{1}{3} (X_1 X_2-X_2^2)^2 
+ \frac{5}{9} (X_1 X_2)^2 + \frac{395}{1764} X_2^4 \\
& + (2 X_1^2+ X_1 X_2- \frac{8}{3} X_2^2)^2 
+ (\frac{4}{3} X_1 X_2+ \frac{3}{2}  X_2^2)^2+(\frac{2}{7} X_2^2)^2) \,.
\end{align*}
\end{example}
\subsection{Correctness and bit size of the output}
\label{sec:bitsize}
%
% Let us first prove that this is sufficient to consider an integer radius $R$ of bit size upper bounded by $\bigo{(\tau d^{?n})}$. 
%
Let $f \in \Z[X]\cap \mathring{\Sigma}[X]$ of degree $d = 2k$,
$\tau := \tau(f)$ and $Q := \polytope(f)/2 \cap \N^n$.
\begin{proposition}
\label{th:boundR}
Let $G$ be a positive definite Gram matrix associated to $f$ and {take
$0 < \varepsilon \in \Q$ as in Proposition \ref{th:boundeps} so that} 
$f_\varepsilon = f - \varepsilon \sum_{\alpha \in Q} X^{2
  \alpha}\in \mathring{\Sigma}[X]$.
Then, there exist positive integers $\delta$, $R$ such that $G - \varepsilon I$ is a Gram matrix associated to $f_\varepsilon$, satisfies
$G - \varepsilon I \succeq 2^{- \delta} I$ and
{$\sqrt{\trace{(({G-\varepsilon I})^2)}} \leq R$}. Also, the maximal
bit sizes of $\delta$ and $R$ are upper bounded by
\revise {$\tau  d^{d^{\bigo{ (n)  }}}$}.
%{$\bigo{( \tau \cdot (4 d+2)^{3n+3} )}$ and $\bigo{( \tau \cdot (4 d+2)^{4n+3} )}$, respectively}.
\end{proposition}
\revise{
\begin{proof}
  Let $\lambda = \lambda_{\min} (G)$ be the smallest eigenvalue of $G$. By
  Proposition~\ref{th:boundeps}, $G \succeq \varepsilon I$ for
  $\varepsilon = \frac{1}{2^N} \leq \frac{\lambda}{2}$ with $N \leq \tau  d^{d^{\bigo{ (n)  }}}$. 
  {By defining 
  $\delta := N+1$}, 
  $2^{-\delta} = \frac{1}{2^{N+1}} \leq \frac{\lambda}{4} <
  \frac{\lambda}{2}$,
  yielding
  $G - \varepsilon \succeq \frac{\lambda}{2} I \succeq 2^{- \delta}I$.
{  As $N \leq \tau  d^{d^{\bigo{ (n)  }}}$, one has $\delta \leq \tau  d^{d^{\bigo{ (n)  }}}$.}
The bound on $R$ follows directly from Proposition~\ref{th:boundGram}.
\end{proof}}
\if{
\begin{proof}
  Let $\lambda$ be the smallest eigenvalue of $G$. By
  Proposition~\ref{th:boundeps}, $G \succeq \varepsilon I$ for
  $\varepsilon = \frac{1}{2^N} \leq \frac{\lambda}{2}$ with $N \leq \bigo{( \tau \cdot (4 d+2)^{3n+3} )}$. 
  {By defining 
  $\delta := N+1$}, 
  $2^{-\delta} = \frac{1}{2^{N+1}} \leq \frac{\lambda}{4} <
  \frac{\lambda}{2}$,
  yielding
  $G - \varepsilon \succeq \frac{\lambda}{2} I \succeq 2^{- \delta}I$.
{  As $N \leq \bigo{( \tau \cdot (4 d+2)^{3n+3} )}$, one has $\delta \leq \bigo{( \tau \cdot (4 d+2)^{3n+3} )}$.}

As in the proof of Proposition~\ref{th:boundeps}, we consider the largest
eigenvalue $\lambda'$ of the Gram matrix $G$ of $f$ and prove that the set $\{e'
\in \R : \forall \x \in \R^n, - f(\x) + e' \sum_{\alpha \in Q} \x^{2 \alpha}
\geq 0\}$ is not empty. {We apply again Proposition~\ref{prop:ori-cpm} as in the
  proof of Proposition~\ref{th:boundeps} to establish that this set contains an
  interval $]0, \frac{1}{2^N}[$ with $N \leq \bigo{( \tau \cdot (4 d+2)^{3n+3}
    )}$. This allows in turn to obtain a rational upper bound $\varepsilon'$ of
  $\lambda'$ with bit size $\bigo{( \tau \cdot (4 d+2)^{3n+3} )})$.} The size of
$G$ is bounded by $\binom{n+k}{n}$, thus the trace of $G^2$ is less than
$\binom{n+k}{n} \varepsilon'^2$. Using that for all $k \geq 2$,
\[
\binom{n+k}{n} = \frac{(n+k) \cdots (k+1)}{n!} =
(1+\frac{k}{n})(1+\frac{k}{n-1})\cdots(1+k)     \leq k^{n-1} (1+k) \leq 2 k^n \leq d^n \,,
\]
one has
{$\sqrt{\trace{(({G-\varepsilon I})^2})} \leq \sqrt{\trace{(G^2)}} \leq d^{\frac{n}{2}}
  \varepsilon' = \bigo{( \tau \cdot (4 d+2)^{4n+3} )} $}.
\end{proof}
}\fi
\begin{proposition}
\label{th:bitmultivsos}
Let $f$ be as above. When applying Algorithm~$\intsos$ to $f$, the procedure
always terminates and outputs a weighted SOS decompositon of $f$ with rational
coefficients. The maximum bit size of the coefficients involved in this SOS
decomposition is upper bounded by \revise {$\tau  d^{d^{\bigo{ (n)  }}}$}.
%{$\bigo{( \tau \cdot (4 d+2)^{4n+3} )}$}.
\end{proposition}
\begin{proof}
  Let us first consider the loop of Algorithm~$\intsos$ defined from
  line~\lineref{line:epsi} to line~\lineref{line:epsf}.  From
  Proposition~\ref{th:boundeps}, this loop terminates when
  $f_\varepsilon \in \mathring{\Sigma}[X]$ for
  $\varepsilon = \frac{1}{2^N}$ and 
\revise {$N \leq \tau  d^{d^{\bigo{ (n)  }}}$}. 
%  {$N \leq \bigo{( \tau \cdot (4 d+2)^{3n+3} )}$}.
 
  When calling the $\sdp$ function at line~\lineref{line:sdp} to solve
  SDP~\eqref{eq:dualsdp} with precision parameters $\delta$ and $R$,
  we compute an approximate Gram matrix $\tilde{G}$ of $f_\varepsilon$
  such that $\tilde{G} \succeq 2^{\delta} I $ and
  $\trace{(\tilde{G}^2)} \leq R^2$.  From
  Proposition~\ref{th:boundR}, this procedure succeeds for large
  enough values of $\delta$ and $R$ of bit size upper bounded by
%{$\bigo{( \tau \cdot (4 d+2)^{4n+3} )}$}. 
\revise {$\tau  d^{d^{\bigo{ (n)  }}}$}. 
In this case, we obtain a positive rational
  approximation $\tilde{\lambda} \geq 2^{-\delta}$ of the smallest
  eigenvalue of $\tilde{G}$.

  Then the Cholesky decomposition of $\tilde{G}$ is computed when
  calling the $\cholesky$ function at line~\lineref{line:chol}.  The
  decomposition is guaranteed to succeed by selecting a large enough
  $\delta_c$ such that~\eqref{eq:chol2} holds. Let $r$ be the size of
  $\tilde{G}$ and $\delta_c$ be the smallest integer such that
  $2^{-\delta_c} < \frac{2^{-\delta}}{r^2 + r + (r-1) 2^{-\delta}}$.
  Since the function $x \mapsto \frac{x}{r^2 + r + (r-1) x}$ is
  increasing on $[0,\infty)$ and
  $\tilde{\lambda} \geq 2^{-\delta}$,~\eqref{eq:chol2} holds. We
  obtain an approximate weighted SOS decomposition
  $\sum_{i=1}^r s_i^2$ of $f_\varepsilon$ with rational coefficients.

  Let us now consider the remainder polynomial
  $u = f_\varepsilon - \sum_{i=1}^r s_i^2$. The second loop of
  Algorithm~$\intsos$ defined from line~\lineref{line:deltai} to
  line~\lineref{line:deltaf} terminates when for all $\alpha \in Q$,
  $\varepsilon_\alpha \geq 0$.  This condition is fulfilled when for
  all $\alpha \in Q$,
  $\varepsilon- \sum_{\beta \in Q} |u_{\alpha + \beta}|/2 + u_\alpha
  \geq 0$.
  This latter condition holds when for all
  $\gamma \in \spt{u}$, $|u_\gamma| \leq \frac{\varepsilon}{r}$.
%Since the cardinal $r$ of $Q$ is upper bounded by $\binom{n+k}{n}$, this latter condition is fulfilled when for all $\gamma \in \spt{u}$, $|u_\gamma| \leq \frac{\varepsilon}{\binom{n+k}{n}}$.

  Next, we show that this happens when the precisions $\delta$ of 
  $\sdp$ and $\delta_c$ of $\cholesky$ are both large enough.  From the definition
  of $u$, one has for all $\gamma \in \spt{u}$,
  $u_\gamma = f_\gamma - \varepsilon_{\gamma} - (\sum_{i=1}^r
  s_i^2)_\gamma$,
  where $\varepsilon_\gamma = \varepsilon$ when $\gamma \in (2\N)^n$
  and $\varepsilon_\gamma = 0$ otherwise.  The positive definite
  matrix $\tilde{G}$ computed by the SDP solver is an approximation of
  an exact Gram matrix of $f_\varepsilon$.  At precision $\delta$, one
  has for all $\gamma \in \spt{f}$,
  $\tilde{G} \succeq 2^{- \delta} I $ and 
\[
|f_\gamma - \varepsilon_\gamma - \trace{( \tilde{G} B_\gamma)} | = |f_\gamma - \varepsilon_\gamma - \sum_{\alpha + \beta = \gamma}
  \tilde{G}_{\alpha, \beta}|
  \leq 2^{-\delta} \,.
  \]
 % or equivalently
%  $|f_\gamma - \varepsilon_\gamma - \sum_{\alpha + \beta = \gamma}
%  \tilde{G}_{\alpha, \beta}| \leq 2^{-\delta}$.
%

In addition, it follows from~\eqref{eq:chol1} that the approximated
  Cholesky decomposition $L L^T$ of $\tilde{G}$ performed at precision
  $\delta$ satisfies $L L^T = \tilde{G} + F$ with
\[
|F_{\alpha,\beta} | \leq \frac{(r+1)2^{-\delta_c}}{1 -
    (r+1)2^{-\delta_c}} |\tilde{G}_{\alpha,\alpha} \,
  \tilde{G}_{\beta,\beta}|^{\frac{1}{2}} \,,
  \]
  for all $\alpha,\beta \in Q$.
Moreover, by using Cauchy-Schwartz inequality, one has
\[\sum_{\alpha \in Q} \tilde{G}_{\alpha,\alpha} = \trace{\tilde{G}}
\leq \sqrt{\trace{I}} \sqrt{\trace{\tilde{G}^2}} \leq \sqrt{r} R \,.\]
For all $\gamma \in \spt{u}$, this yields
\[\bigl \lvert \sum_{\alpha + \beta = \gamma} \tilde{G}_{\alpha,\alpha}
\, \tilde{G}_{\beta,\beta}\bigr \rvert^{\frac{1}{2}}\leq \sum_{\alpha
  + \beta = \gamma} \frac{ \tilde{G}_{\alpha,\alpha} +
  \tilde{G}_{\beta,\beta} }{2} \leq \trace{\tilde{G}} \leq \sqrt{r} R \,,
\]
where the first inequality comes again from Cauchy-Schwartz
inequality.

Thus, for all $\gamma \in \spt{u}$, one has
\[
\bigl \lvert \sum_{\alpha + \beta = \gamma} \tilde{G}_{\alpha, \beta}
- (\sum_{i=1}^r s_i^2)_\gamma \bigr \rvert = \bigl \lvert \sum_{\alpha
  + \beta = \gamma} \tilde{G}_{\alpha, \beta} - \sum_{\alpha + \beta =
  \gamma} (L L^T)_{\alpha, \beta} \bigr \rvert = \bigl \lvert
\sum_{\alpha + \beta = \gamma} F_{\alpha,\beta} \bigr \rvert \,, \]
which is bounded by
\[ \frac{(r+1)2^{-\delta_c}}{1 - (r+1)2^{-\delta_c}} \sum_{\alpha +
  \beta = \gamma} |\tilde{G}_{\alpha,\alpha} \,
\tilde{G}_{\beta,\beta}|^{\frac{1}{2}} \leq
\frac{\sqrt{r}(r+1)2^{-\delta_c} \, R}{1 - (r+1)2^{-\delta_c}} \,.\]
Now, let us take the smallest $\delta$ such that
$2^{-\delta} \leq \frac{\varepsilon}{2 r} = \frac{1}{2^{N+1} r}$ as
well as the smallest $\delta_c$ such that
$\frac{\sqrt{r}(r+1)2^{-\delta_c} \, R}{1 - (r+1)2^{-\delta_c}} \leq
\frac{\varepsilon}{2 r}$,
that is $\delta = \lceil N + 1 + \log_2 r \rceil$ and
$\delta_c = \lceil \log_2 R + \log_2 (r+1) + \log_2 (2^{N+1} r \sqrt{r} + 1)
\rceil$.

%TODO Note that choosing $\delta_c$ as above does not compromise the fact that~\eqref{eq:chol2} holds.
From the previous inequalities, for all
$\gamma \in \spt{u}$, it holds that
\[
|u_\gamma| = | f_\gamma - \varepsilon_{\gamma} - (\sum_{i=1}^r
s_i^2)_\gamma| \leq |f_\gamma - \varepsilon_\gamma - \sum_{\alpha +
  \beta = \gamma} \tilde{G}_{\alpha, \beta}| + |\sum_{\alpha + \beta =
  \gamma} \tilde{G}_{\alpha, \beta} - (\sum_{i=1}^r s_i^2)_\gamma |
\leq \frac{\varepsilon}{2 r} + \frac{\varepsilon}{2 r} =
\frac{\varepsilon}{r} \,. \]
This ensures that Algorithm~$\intsos$ terminates. 

Let us note
\[\Delta(u) := \{(\alpha, \beta) : \alpha + \beta \in \spt{u} \,,
\alpha,\beta \in Q\,, \alpha \neq \beta \} \,. \]
When terminating, the first output $\clist$ of Algorithm~$\intsos$ is a 
list of non-negative rational numbers containing the list
$[1, \dots, 1]$ of length $r$, the list
$\bigl\{\frac{|u_{\alpha + \beta}|}{2} : (\alpha, \beta) \in \Delta(u)
\bigr\}$
and the list $\{ \varepsilon_\alpha : \alpha \in Q \}$.  The
second output $\slist$ of Algorithm~$\intsos$ is a list of polynomials 
containing the list $[s_1, \dots, s_r]$, the list
$\{ X^\alpha + \sgn{(u_{\alpha+\beta})} X^\beta : (\alpha, \beta) \in
\Delta(u) \}$
and the list $\{ X^\alpha : \alpha \in Q \}$.  From the output, we
obtain the following weigthed SOS decomposition
\[
f = \sum_{i=1}^r s_i^2 + \sum_{\mbox{\tiny
    $(\alpha, \beta) \in \Delta(u)$ }} \dfrac{|u_{\alpha + \beta}|}{2}
(X^\alpha + \sgn{(u_{\alpha+\beta})} X^\beta)^2 + \sum_{\mbox{\tiny
    $\alpha \in Q$ }} \varepsilon_\alpha X^{2 \alpha} \,. \]
Now, we bound the bit size of the coefficients. %involved in this
%decomposition.  
\if{
{Since $r \leq \binom{n+k}{n} \leq d^n$ and
$N \leq \bigo{( \tau \cdot (4 d+2)^{3n+3} )}$, one has $\delta \leq  \bigo{( \tau \cdot (4 d+2)^{3n+3} )}$. Similarly, $\delta_c \leq \bigo{( \tau \cdot (4 d+2)^{4n+3} )}$.}
}\fi

\revise {
Since $r \leq \binom{n+k}{n} \leq d^n$ and 
$N \leq \tau  d^{d^{\bigo{ (n)  }}}$, one has $\delta \leq  \tau  d^{d^{\bigo{ (n)  }}}$.
Similarly, $\delta_c \leq \tau  d^{d^{\bigo{ (n)  }}}$.
}  

%Similarly, $R \leq \tau d^{\bigo{(n)}}$
%implies the same upper bound for $\delta_c$.
%
This bounds also the maximal bit size of the coefficients involved in
the approximate decomposition $\sum_{i=1}^r s_i^2$ as well as the coefficients of $u$. In the worst case, the coefficient
$\varepsilon_\alpha$ involved in the exact SOS decomposition is equal
to
$\varepsilon- \sum_{\beta \in Q} |u_{\alpha + \beta}|/2 + u_\alpha$
for some $\alpha \in Q$. Using again that the cardinal $r$ of $Q$ is less than $\binom{n+k}{n} \leq d^n$, we obtain a
maximum bit size upper bounded by \revise{$\tau  d^{d^{\bigo{ (n)  }}}$}.
%
%Overall, the maximum bit size of the coefficients involved in the
%output of Algorithm~$\intsos$ is upper bounded by
%$\tau d^{\bigo{(n)}}$.
\end{proof}
\subsection{Bit complexity analysis}
\label{sec:bitop}
\begin{theorem}
\label{th:costmultivsos}
\if{
For $f$ as above, there exist $\varepsilon$, $\delta$, $R$, $\delta_c$ of bit sizes {upper bounded by $\bigo{( \tau \cdot (4 d+2)^{4n+3} )}$ such that $\intsosfun{f}{\varepsilon}{\delta}{R}{\delta_c}$ runs in
boolean time $\bigo{( \tau^2 \cdot (4 d+2)^{15 n+6} )}$}.
}\fi
\revise{
For $f$ as above, there exist $\varepsilon$, $\delta$, $R$, $\delta_c$ of bit sizes {upper bounded by $\tau  d^{d^{\bigo{ (n)  }}}$ such that $\intsosfun{f}{\varepsilon}{\delta}{R}{\delta_c}$ runs in
boolean time $\tau^2  d^{d^{\bigo{ (n)  }}}$}.
}
\end{theorem}
\begin{proof}
  We consider $\varepsilon$, $\delta$, $R$ and $\delta_c$ as in the
  proof of Proposition~\ref{th:bitmultivsos}, so that
  Algorithm~$\intsos$ only performs a single iteration within the two
  while loops before terminating. Thus, the bit size of each input
  parameter is upper bounded by 
 % {$\bigo{( \tau \cdot (4 d+2)^{4n+3} )}$}.
 \revise {$\tau  d^{d^{\bigo{ (n)  }}}$}.

Computing $\polytope(f)$ with 
the  quickhull algorithm runs in boolean time $\bigo{(V^2)}$  for
  a polytope with $V$ vertices. In our case $V \leq \binom{n+d}{n} \leq 2 d^n$,
  so that this procedure runs in boolean time
 $\bigo{(d^{2 n})}$.
  Next, we investigate the computational cost of the call to $\sdp$ at
  line~\lineref{line:sdp}.  Let us note $\nsdp = r$ (resp.~$\msdp$)
  the size (resp.~number of entries) of $\tilde{G}$.  This step
  consists of solving SDP~\eqref{eq:dualsdp}, which is performed in
  $\bigo{( \nsdp^4 \log_2 (2^\tau \nsdp \, R \, 2^\delta) )}$ iterations
  of the ellipsoid method, where each iteration requires
  $\bigo{(\nsdp^2(\msdp+\nsdp) )}$ arithmetic operations over
  $\log_2 (2^\tau \nsdp \, R \, 2^\delta)$-bit numbers (see e.g.~\cite{GroetschelLovaszSchrijver93,porkolab1997complexity}). Since
  $\msdp, \nsdp $ are both bounded above by $\binom{n+d}{n} \leq 2 d^n$, one has 
  \revise{
\begin{align*}
\log_2 (2^\tau \nsdp \, R \, 2^\delta) & \leq \tau  d^{d^{\bigo{ (n)  }}}  \,,  \\
\nsdp^2(\msdp+\nsdp) & \leq  \bigo{( d^{3 n})} \,, \\
\nsdp^4 \log_2(2^\tau \nsdp \, R \, 2^\delta) & \leq  \tau  d^{d^{\bigo{ (n)  }}} \,.  
\end{align*}
  Overall, the ellipsoid algorithm runs in boolean time
  $\tau^2  d^{d^{\bigo{ (n)  }}}$ to compute the approximate Gram matrix
  $\tilde{G}$.
  We end with the cost of the call to $\cholesky$ at
  line~\lineref{line:chol}. Cholesky's decomposition is performed in
  $\bigo{(\nsdp^3)}$ arithmetic operations over $\delta_c$-bit
  numbers. {Since $\delta_c \leq \tau  d^{d^{\bigo{ (n)  }}}$, the function
  runs in boolean time $\tau  d^{d^{\bigo{ (n)  }}}$.}  The other elementary
  arithmetic operations performed while running Algorithm~$\intsos$
  have a negligible cost w.r.t.~to the $\sdp$ procedure.   
  }
  \if{
\begin{align*}
\log_2 (2^\tau \nsdp \, R \, 2^\delta) & \leq  \bigo{( \tau \cdot (4 d+2)^{4n+3} )}  \,,  \\
\nsdp^2(\msdp+\nsdp) & \leq  \bigo{( d^{3 n})} \,, \\
\nsdp^4 \log_2(2^\tau \nsdp \, R \, 2^\delta) & \leq  \bigo{( \tau \cdot (4 d+2)^{8n+3} )} \,.  
\end{align*}
  Overall, the ellipsoid algorithm runs in boolean time
  $\bigo{( \tau^2 \cdot (4 d+2)^{15 n+6} )}$ to compute the approximate Gram matrix
  $\tilde{G}$.
  We end with the cost of the call to $\cholesky$ at
  line~\lineref{line:chol}. Cholesky's decomposition is performed in
  $\bigo{(\nsdp^3)}$ arithmetic operations over $\delta_c$-bit
  numbers. {Since $\delta_c \leq \bigo{( \tau \cdot (4 d+2)^{4n+3} )}$, the function
  runs in boolean time $\bigo{( \tau \cdot (4 d+2)^{7n+3} )}$.}  The other elementary
  arithmetic operations performed while running Algorithm~$\intsos$
  have a negligible cost w.r.t.~to the $\sdp$ procedure. 
}  \fi
  %OLD The total boolean running time is bounded by $\tau^2 d^{\bigo{(n)}}$.
  %, yielding
  %the desired result.
\end{proof}

{
\subsection{Comparison with the rounding-projection algorithm of Peyrl and Parrilo}
\label{sec:pp}
}
{
We recall the algorithm {designed in~\cite{PaPe08}}. We denote this rounding-projection algorithm by~$\PP$.}

\begin{algorithm}
\caption{{$\PP$}%: algorithm to compute weighted SOS decompositions of multivariate polynomials in $\mathring{\Sigma}[X]$.
}
\label{alg:PP}
{
\begin{algorithmic}[1]
%\Require $f \in \mathring{\Sigma}_\Z[X]$, positive $\varepsilon \in \Q$, precision parameters $\delta, R \in \N$ for the SDP solver 
\Require $f \in \Z[X]$, rounding precision $\delta_i \in \N$, precision parameters $\delta, R \in \N$ for the SDP solver
\Ensure list $\clist$ of numbers in $\Q$ and list $\slist$ of polynomials in $\Q[X]$
\State $P := \polytope{(f)}$, $Q := P/2 \cap \N^n$
 \label{line:npPP}
\State ok := false
\While {not ok} \label{line:PPi}
\State $(\tilde{G}, \tilde{\lambda}) \gets \sdpfun{f}{\delta}{R}$ \label{line:sdpPP}
\State $G' \gets \roundfun{\tilde{G}}{\delta_i}$ \label{line:roundPP}
\For {$\alpha, \beta \in Q$}  
\State $\eta(\alpha+\beta) \gets \# \{(\alpha',\beta') \in Q^2 \mid \alpha'+\beta' = \alpha + \beta \}$
\State $G(\alpha,\beta) := G'(\alpha,\beta) - \frac{1}{\eta(\alpha+\beta)} \Bigl( \sum_{\alpha'+\beta'=\alpha + \beta} G'(\alpha',\beta')-f_{\alpha+\beta} \Bigr)$ \label{line:projPP}
\EndFor
\State $(c_1,\dots, c_r, s_1,\dots,s_r) \gets \ldlfun{G}$ \label{line:cholPP} \Comment{$f =  \sum_{i=1}^r c_i s_i^2$}
\If {$c_1,\dots,c_r \in \Q^{>0}, s_1,\dots,s_r \in \Q[X]$} ok := true \label{line:LDLok}
\Else $\ \delta \gets 2 \delta$, $R \gets 2 R$, $\delta_c \gets 2 \delta_c$
\EndIf
\EndWhile \label{line:PPf}
\State $\clist \gets [c_1,\dots,c_r] $, $\slist \gets [s_1,\dots,s_r]$
\State \Return $\clist$, $\slist$
\end{algorithmic}
}
\end{algorithm}

{
The first main step in Line~\lineref{line:roundPP} consists of rounding the approximation $\tilde{G}$ of a Gram matrix associated to $f$ {up to precision $\delta_i$}. 
%to obtain a matrix $G'$ with rational entries. 
The second main step in Line~\lineref{line:projPP} consists of computing the
orthogonal projection $G$ of $G'$ on an adequate affine subspace in such a way
that $\sum_{\alpha + \beta = \gamma} G_{\alpha,\beta} = f_\gamma$, for all
$\gamma \in \spt{f}$. For more details on this orthogonal projection, we refer
to~\cite[Proposition~7]{PaPe08}. The algorithm then performs
in~\eqref{line:cholPP} an exact diagonalization of the matrix $G$ via the $L D
L^T$ decomposition (see e.g.~\cite[\S~4.1]{Golub96}). It is proved
in~\cite[Proposition~8]{PaPe08} that for $f \in \mathring{\Sigma}[X]$,
Algorithm~$\PP$ returns a weighted SOS decomposition of $f$ with rational
coefficients when the precision of the rounding and SDP solving steps are large enough.}

{The main differences w.r.t.~Algorithm~$\intsos$ are that $\PP$ does not perform a perturbation of the input polynomial $f$ and computes an exact $L D L^T$ decomposition of a Gram matrix $G$. In our case, we compute an approximate Cholesky's decomposition of $\tilde{G}$ instead of a projection, then perform an exact compensation of the error terms,  thanks to the initial perturbation.}

%{The next theorem Even though both algorithms have the same exponential bit complexity,~$\PP$ returns SOS decomposition with coefficients of bit size upper bounded: 
{The next result gives upper bounds on the bit size of the coefficients involved in the SOS decomposition returned   by~$\PP$ as well as upper bounds on the boolean running time.
It turns out that $\intsos$ and $\PP$ have the same exponential bit complexity. 
%, the upper bound estimates are larger in the case of~$\PP$. 
It would be worth investigating whether these bounds are tight in general.
}
\begin{theorem}
\label{th:bitPP}
For $f$ as above, there exist $\delta_i$, $\delta$, $R$ of bit sizes $\leq \tau \cdot d^{d^{\bigo{ (n)  }}}$ such that
$\PPfun{f}{\delta_i}{\delta}{R}$ outputs a rational SOS decomposition of $f$
with rational coefficients. The maximum bit size of the coefficients involved in this SOS decomposition is upper bounded by \revise{$\tau  d^{d^{\bigo{ (n)  }}}$} and the  
boolean running time is \revise{$\tau^2  d^{d^{\bigo{ (n)  }}}$}.
\end{theorem}
\begin{proof}
Let us assume that Algorithm~$\PP$ returns a matrix $G \succ 0$ associated to $f$ with smallest eigenvalue $\lambda$ and let $N \in \N$ be the smallest integer such that $2^{-N} \leq \lambda$. As in Proposition~\ref{th:boundR}, one proves that the bit size of $N$ is upper bounded by $\tau  d^{d^{\bigo{ (n)  }}}$.
%As in Proposition~\ref{th:boundR}, the size of $N$ and $R$ are upper bounded by $\tau d^{\bigo{(n)}}$.
%
By~\cite[Proposition~8]{PaPe08}, Algorithm~$\PP$ terminates and outputs such a matrix $G$ together with a weighted rational SOS decomposition of $f$ if $2^{-\delta_i} + 2^{-\delta'} \leq 2^{-N}$, where $\delta'$ stands for the euclidean distance between $G'$ and $G$, yielding 
\[
\sqrt{\sum_{\alpha,\beta \in Q} (G_{\alpha,\beta} - G'_{\alpha,\beta})^2 } = 2^{-\delta'} \,.
\]
For all $\alpha,\beta \in Q$, one has $\vert G'_{\alpha,\beta} - \tilde{G}_{\alpha,\beta} \vert \leq 2^{-\delta_i}$. 
As in the proof of Proposition~\ref{th:bitmultivsos}, at SDP precision $\delta$, one has for all $\gamma \in \spt{f}$, $\tilde{G} \succeq 2^{-\delta} I$ and 
\[
\vert f_\gamma - \sum_{\alpha + \beta = \gamma} \tilde{G}_{\alpha,\beta} \vert \leq 2^{-\delta} \,.
\]
For all $\alpha, \beta \in Q$,  let us define $e_{\alpha,\beta} := \sum_{\alpha'+\beta'=\alpha + \beta} G'(\alpha',\beta')-f_{\alpha+\beta}$ and note that
\[
\vert e_{\alpha,\beta} \vert \leq 
 \sum_{\alpha'+\beta'=\alpha + \beta} \Big\vert G'(\alpha',\beta')-  \tilde{G}(\alpha',\beta') \Big\vert + 
\Big\vert \sum_{\alpha'+\beta'=\alpha + \beta} \tilde{G} (\alpha',\beta') - f_{\alpha+\beta} \Big\vert \leq \eta(\alpha + \beta) 2^{-\delta_i} + 2^{-\delta} \,.
\]
For all $\alpha,\beta \in Q$, we use the fact that $\eta(\alpha + \beta) \geq 1$ and that the cardinal of $Q$ is less than the size $r$ of $G$, with $r \leq d^n$, to obtain
\[
2^{-\delta'} = \sum_{\alpha,\beta \in Q} \frac{e_{\alpha,\beta}}{\eta(\alpha + \beta)} \leq d^{2n} (2^{-\delta_i} + 2^{-\delta}).
\]
To ensure that $2^{-\delta_i} + 2^{-\delta'} \leq 2^{-N}$, it is sufficient to have $(d^{2n} + 1) 2^{-\delta_i} + d^{2n} 2^{-\delta} \leq 2^{-N}$, which is obtained with $\delta_i$ and $\delta$ with bit size upper bounded by $\tau  d^{d^{\bigo{ (n)  }}}$. The bit size of the coefficients involved in the weighted SOS decomposition is upper bounded by the output bit size of the $LDL^T$ decomposition of the matrix $G$, that is $\bigo{(\delta_i r^3)} = \tau  d^{d^{\bigo{ (n)  }}}$.\\
The bound on the running time is obtained exactly as in Theorem~\ref{th:bitmultivsos}.
\end{proof}
%%% Local Variables:
%%% mode: latex
%%% TeX-master: "multivsos"
%%% End:

%!TEX root = multivsosJSC.tex
\section{Exact Reznick and Hilbert-Artin's representations}
\label{sec:reznickhilbert}
Next, we show how to apply Algorithm~$\intsos$ to decompose positive definite forms {and positive polynomials} into SOS of rational functions. 

\subsection{{Exact Reznick's representations}}
\label{sec:reznick}

Let $G_n := \sum_{i=1}^n X_i^2$ and
$\Sb^{n-1} := \{\bmx \in \R^n : G_n(\bmx) = 1 \}$ be the unit
$(n-1)$-sphere.
A positive definite form $f\in \R[X]$ is a homogeneous polynomial
which is positive over $\Sb^{n-1}$. For such a form, we set
\[
  \varepsilon(f) := \frac{\min_{\bmx \in \Sb^{n-1}} f(\bmx)}{\max_{\bmx
    \in \Sb^{n-1}} f(\bmx)},
\]
which measures how close $f$ is to having a zero in
$\Sb^{n-1}$. While there is no guarantee that $f \in \Sigma[X]$,~\cite{Reznick95} proved that for large enough $D \in \N$, $f G_n^D \in \Sigma[X]$. 
%The proof being based on prior work by Polya~\cite{Polya}, 
Such SOS decompositions are called {\em Reznick's representations} and $D$ is called the {\em Reznick's degree}.
%
% We first recall the following result.
% %
% \begin{theorem}~\cite[Theorem~3.12]{Reznick95}
% \label{th:polya}
% Let $f$ be a positive definite form of degree $d$ in $\Z[X]$ and
% $D \geq \frac{n d (d-1)}{4 \log 2 \, \varepsilon(f)} - \frac{n +
%   d}{2}$.  Then $f \, G_n^D \in {\Sigma}_\Z[X]$.
% \end{theorem}
%
The next result states that for large enough $D \in \N$, $f G_n^D \in \mathring{\Sigma}[X]$, as a direct consequence of~\cite{Reznick95}.
\begin{lemma}
\label{th:reznickint}
Let $f$ be a positive definite form of degree $d = 2 k$ in $\Z[X]$ and
$D \geq \frac{n d (d-1)}{4 \log 2 \, \varepsilon(f)} - \frac{n +
  d}{2} {+ 1}$.  Then $f \, G_n^D \in \mathring{\Sigma}[X]$.
\end{lemma}
\begin{proof}
\if{
By~\cite[Theorem~3.12]{Reznick95}, for any positive integer $D \geq \frac{n d (d-1)}{4 \log 2 \, \varepsilon(f)} - \frac{n +
  d}{2}$, there exists a form $\sigma \in \Sigma[X]$ such that $\sigma = f \, G_n^D$.
Let us assume that $\sigma (\bmx) = 0$ for some $\bmx \in \R^n$. 
Then one has either $G_n^D (\bmx) = 0$ or $f(\bmx) = 0$.
In the former case, this implies that $\bmx=0$.
In the latter case, this also implies that $\bmx=0$ since $f$ is positive definite. 
Then the form $\sigma$ is positive definite, which yields the desired result.
}\fi
%\if{
{
%
%  Let $P := \polytope{(f)}$, $Q := P/2 \cap \N^n$ and
%  $t := \sum_{\alpha \in Q} X^{2 \alpha}$. Since $f$ is a form, then
%  each term $X^{2 \alpha}$ has degree $d = 2k$, for all $\alpha \in Q$,
%  thus $t$ is a form.  
  {First, for any positive 
  $e < \min_{\bmx \in \Sb^{n-1}} f(\bmx)$, the form $(f - e G_n^k)$ is positive  on $\Sb^{n-1}$.
 % Indeed, if $f(\bmx) - e G_n(\bmx)^{d} = 0$, then it implies that f(\bmx)
  Then, for any nonzero
  $\bmx \in \R^n$, one has
  \[f(\bmx) - e G_n(\bmx)^k = G_n(\bmx)^k \Big(
  f\Bigl(\frac{\bmx}{G_n(\bmx)}\Bigr) - e
   \Big) > 0 \,,\]
  implying that $(f - e G_n^k)$ is positive definite.}  Next,
  \cite[Theorem~3.12]{Reznick95} implies that for any positive integer $D_e$ such that
\[
D_e \geq \underline{D_e} := \frac{n d (d-1)}{4 \log 2 \,
    \varepsilon(f - e G_n^k)} - \frac{n + d}{2}
\,,    \]
  one has $(f - e G_n^k) \, G_n ^{D_e} \in \Sigma[X]$. 
{
%Let $Q' := \{\alpha+\beta :  \alpha \in Q, |\alpha + \beta| = k + D_e \}$.
One has $G_n ^{k + D_e} = \sum_{|\alpha| = k + D_e} \binom{\newjsc{k+}D_e}{\alpha_1! \dots \alpha_n!} X^{2 \alpha}$.
Let $v_{k + D_e}(X)$ be the vector of monomials with exponents in $\N_{k+D_e}^{n}$. 
Then, one can write $G_n ^{k+D_e} = v_{k + D_e}^T A v_{k + D_e}$ with $A$ being a diagonal matrix  with positive entries $\binom{\newjsc{k+}D_e}{\alpha_1! \dots \alpha_n!}$, thus $A \succ 0$. 
Next we select $e$ small enough so that there is no term cancellation in $(f - e G_n^k)$, ensuring that the \newjsc{Newton polytope  of $(f - e G_n^k)$ is equal to $\N_{2 k}^{n}$.
This in turn implies that the Newton polytope of $(f - e G_n^k) \, G_n ^{D_e}$ is equal to $\N_{2( k+D_e)}^{n}$.}
%By definition of the Newton polytope, the support of $f$ is included in the support of $t$, thus the support of $f G_n ^{D_e}$ is included in the support of $t G_n ^{D_e}$.
%Hence, one has $\mathcal{C}(f G_n^{D_e} - e t G_n^{D_e})/2 \cap \N^n \subseteq Q'$.
Since $(f - e G_n^k) \, G_n ^{D_e} \in \Sigma[X]$,  there exists $A' \succeq 0$ indexed by $\N_{k+D_e}^n$ such that $f G_n^{D_e} - e  G_n^{k+D_e} = v_{k + D_e}^T A' v_{k + D_e}$.
This yields $f G_n^{D_e} = v_{k + D_e}^T (\newjsc{e} A + A') v_{k + D_e}$. 
Since $\newjsc{e}  A + A' \succ 0$, Theorem~\ref{th:intsospdGram} implies that   $f \, G_n ^{D_e} \in \mathring{\Sigma}[X]$.
  }
  
  %As in the proof of Proposition~\ref{th:boundeps}, 
   Next, with 
  $\underline{D} := \frac{n d (d-1)}{4 \log 2 \, \varepsilon(f)} -
  \frac{n + d}{2}$, 
  we prove that there exists a large enough $N \in \N$ such that for
  $e = \frac{\min_{\bmx \in \Sb^{n-1}} f(\bmx)}{N }$,
  $\underline{D_e} \leq \underline{D}+1$.  Since
  $f \, G_n^{D_e} \in \mathring{\Sigma}[X]$ for all
  $D_e \geq \underline{D_e}$, this will yield the desired result. 
  For
  any $\bmx \in \Sb^{n-1}$, one has $G_n(\bmx)^k = 1$, thus
  \[\min_{\bmx \in \Sb^{n-1}} (f(\bmx) - e G_n(\bmx)^k) = \min_{\bmx \in \Sb^{n-1}} f(\bmx) - e  \,, \quad \max_{\bmx \in \Sb^{n-1}} (f(\bmx) - e G_n(\bmx)^k) = \max_{\bmx \in \Sb^{n-1}} f(\bmx) - e   \,.\]
  Hence,
  \[\varepsilon(f - e G_n^k) = \frac{\min_{\bmx \in \Sb^{n-1}} f(\bmx) [1 - 1/N] 
    }{\min_{\bmx \in \Sb^{n-1}}
    f(\bmx) [1/\varepsilon(f) - 1/N]} =  \frac{ \varepsilon(f)(N-1)}{N - \varepsilon(f)} \,.\]
  Therefore, one has
  $\underline{D_e} = \frac{N-\varepsilon(f)}{N-1} \frac{n d (d-1)}{4 \log 2 \,
    \varepsilon(f)} - \frac{n + d}{2}$,
  yielding
  $\underline{D_e} - \underline{D} = \frac{1-\varepsilon(f)}{N-1} \frac{n d
    (d-1)}{4 \log 2 \, \varepsilon(f)}$.
  By choosing
  $N > \lfloor \frac{ (1-\varepsilon(f))  n d (d-1)}{4 \log 2 \, \varepsilon(f)} +1
  \rfloor$,
  one ensures that $\underline{D_e} - \underline{D} \leq 1$, which
  concludes the proof.
}
%}\fi
\end{proof}

%emphasize degree bounds. 

%State the main theorem. 

%\subsection{Algorithm~$\reznicksos$}
%\label{sec:polyasos}
%
%exact weighted rational SOS decompositions for a positive definite form 
%Algorithm~$\polyasos$ takes as input $f \in \Z[X]$. The first loop at line~\lineref{line:Di} finds the smallest $D \in \N$ such that $f \, G_n^D \in \mathring{\Sigma}[X]$, thanks to an oracle relying either on SDP or computer algebra (as for~$\intsos$). Then, $\intsos$ is applied on $f \, G_n^D$. 
Algorithm~$\reznicksos$ takes as input $f \in \Z[X]$, finds the smallest $D \in
\N$ such that $f \, G_n^D \in \mathring{\Sigma}[X]$, thanks to an oracle which
decides if some given polynomial is a positive definite form. 
{Further, we denote
by $\interiorsoscone$ a routine which takes as input $f, G_n$ and $D$ and
returns {\sf true} if and only if $f \, G_n^D \in \mathring{\Sigma}[X]$, else it
returns false. Then, $\intsos$ is applied on $f \, G_n^D$. 
}
{
\begin{algorithm}
\caption{$\reznicksos$%: algorithm to compute exact decompositions of positive definite forms into SOS of rational functions.
}
\label{alg:reznicksos}
\begin{algorithmic}[1]
\Require $f \in \Z[X]$, positive $\varepsilon \in \Q$, precision parameters $\delta, R \in \N$ for the SDP solver,  precision $\delta_c \in \N$ for the Cholesky's decomposition
\Ensure list $\clist$ of numbers in $\Q$ and list $\slist$ of polynomials in $\Q[X]$
\State $D := 0$
\While {{$\interiorsoscone(f \, G_n, D) ={\sf false}$}} \label{line:Di} $D \gets D +1$ 
\EndWhile \label{line:Df}
\State \Return $\intsosfun{f \, G_n^D}{\varepsilon}{\delta}{R}{\delta_c}$
\end{algorithmic}
\end{algorithm}
}
\begin{example}
\label{ex:reznick}
Let us apply $\reznicksos$ on the perturbed Motzkin polynomial
\[
f = (1+2^{-20}) (X_3^6 +  X_1^4 X_2^2 + X_1^2 X_2^4) - 3 X_1^2 X_2^2 X_3^2.
\]
With $D = 1$, one has $f \, G_n = (X_1^2 + X_2^2 + X_3^2) \, f \in \mathring{\Sigma}[X]$ and $\intsos$ yields an SOS decomposition of $f \, G_n$ with $\varepsilon = 2^{-20}$, $\delta = R = 60$, $\delta_c = 10$.
\end{example}
\begin{theorem}
\label{th:reznicksos}
Let $f \in \Z[X]$ be a positive definite form of degree $d$,
coefficients of bit size at most $\tau$. On input $f$,
Algorithm~$\reznicksos$ terminates and outputs a weighted SOS
decomposition for $f$. The maximum bit size of the coefficients
involved in the decomposition and the boolean running time of the
procedure are both upper bounded by \revise{$2^{2^{  \tau^{\bigo{(1)}} \cdot (4 d+6)^{\bigo{(n)}} }} $}.
\end{theorem}
The proof of this result relies on the following technical statement whose proof
is postponed in the Appendix. 

\begin{proposition}
  \label{prop:ori-cpm}
  Let $g\in \Z[X]$ of degree $d$ and let $\tau = \tau(g)$. Assume that the
  algebraic set $V(g)\subset \mathbb{C}^n$ defined by $g = 0$ is smooth. Then, there
  exists a polynomial $w\in \Z[X_1]$ of degree $\leq d^n$ with coefficients of
  bit size $\leq \tau \cdot (4 d+2)^{3n}$ such that its set of real roots
  contains the critical values of the restriction of the projection on the
  $X_1$-axis to $V(g)$.
\end{proposition}

\begin{proof}[Proof of Theorem~\ref{th:reznicksos}]
By Lemma~\ref{th:reznickint}, the while loop from
  line~\lineref{line:Di} to ~\lineref{line:Df} is ensured to terminate
  for a positive integer
  $D \geq \frac{n d (d-1)}{4 \log 2 \, \varepsilon(f)} - \frac{n +
    d}{2}$.
  By Proposition~\ref{th:bitmultivsos}, when applying $\intsos$ to
  $f \, G_n^D$, the procedure always terminates. The outputs are a
  list of non-negative rational numbers $[c_1,\dots,c_r]$ and a list
  of rational polynomials $[s_1,\dots,s_r]$ providing the weighted SOS
  decompositon $ f \, G_n^D = \sum_{i=1}^r c_i s_i^2$. Thus, we obtain
  $ f = \sum_{i=1}^r c_i \frac{s_i^2}{G_n^D}$, yielding the first
  claim.

  Since,
  $(X_1^2 + \dots + X_n^2)^D = \sum_{|\alpha| = D} \frac{D!}{\alpha_1!
    \cdots \alpha_n!} \, X^{2 \alpha}$,
  each coefficient of $G_n^D$ is upper bounded by
  $\sum_{|\alpha| = D} \frac{D!}{\alpha_1! \cdots \alpha_n!} = n^D$.
  Thus $\tau(f \, G_n^D) \leq \tau + D \log n$.  Using again
  Proposition~\ref{th:bitmultivsos}, the maximum bit size of the
  coefficients involved in the weighted SOS decomposition of
  $f \, G_n^D$ is upper bounded by
%  {$\bigo{( (\tau + D \log n) (4 d + 8 D +2)^{3n+3} )}$}.
\revise{$(\tau + D \log n) (d+D)^{(d+D)^{\bigo{(n)}}}$}.
\revise{
  Now, we derive an upper
  bound on $D$. 
  Since $f$ is a positive form of degree $d$, one has
  \[\min_{\bmx \in \Sb^{n-1}} f(\bmx) = \max \{e : \forall \bmx \in
  \R^n, f(\bmx) - e G_n(\bmx)^d \geq 0 \} \,. \]
  
%  One has $\min_{\bmx \in \Sb^{n-1}} f(\bmx) := \min \{e \in %\R^{>0} : f(\bmx) - e = 0 \,, \bmx \in \Sb^{n-1} \} $.\\
  %
  We rely on Proposition~\ref{prop:ori-cpm} to show that
  $\min_{\bmx \in \Sb^{n-1}} f(\bmx) \geq 2^{-  \tau \cdot (4 d+6)^{3n+3}  }$.
For this, let us define the polynomial $g(X,E) := f - E G_n^d$ , 
and consider the algebraic set $V$ defined by:
\[
V := \left\{ (\x,e) \in \R^{n+1} : g(\x,e) = \frac{\partial g }{\partial x_1} = \dots = \frac{\partial g}{\partial x_n} = 0 \right\} \,.
\] 
%The degree of $g$ is also $d+D$.
First, note that the minimum of $f$ on the sphere belongs to the projection of $V$ on the $E$-axis.
Using Proposition~\ref{prop:ori-cpm}, there exists a polynomial in $\Z[E]$ of degree less than $(d+1)^{n+1}$ with coefficients of bit size less than $\tau \cdot (4 d + 6)^{3 n + 3}$ such that its set of real roots contains the projection of $V$ on the $E$-axis. 
By Lemma~\ref{th:cauchyint}, the bit size of the minimum of $f$ on the sphere is upper bounded by $\tau \cdot (4 d + 6)^{3 n + 3}$.

  Similarly, we obtain
  $\max_{\bmx \in \Sb^{n-1}} f(\bmx) \leq 2^{ \tau \cdot (4 d+6)^{3n+3} }$ and
  thus $\frac{1}{\varepsilon(f)} \leq 2^{ 2 \tau \cdot (4 d+6)^{3n+3} }$. Overall, we
  obtain
  \[
  \frac{n d (d-1)}{4 \log 2 \, \varepsilon(f)} - \frac{n + d}{2} +1
  \leq D \leq 2^{  \tau \cdot (4 d+6)^{\bigo{(n)} }  } \,. \]
  This implies that
  \[ (\tau + D \log n) \cdot (d+D)^{(d+D)^{\bigo{(n)}}}
  \leq 2^{2^{  \tau^{\bigo{(1)}} \cdot (4 d+6)^{\bigo{(n)}} }} 
   \,.\]
  From Theorem~\ref{th:costmultivsos}, the  running time is
  upper bounded by $(\tau + D \log n)^2 \cdot (d+D)^{(d+D)^{\bigo{(n)}}}$, which
  ends the proof.}
\end{proof}
{The bit complexity of $\reznicksos$ is polynomial in the Reznick's degree $D$
  of the representation. In all the examples we tackled, this degree was rather
  small as shown in Section~\ref{sec:benchs}. }

\subsection{{Exact Hilbert-Artin's representations}}
\label{sec:hilbert}
{
Here, we focus on the subclass of non-negative polynomials in $\Z[X]$ which admit an Hilbert-Artin's representation of the form $f = \frac{\mathring{\sigma}}{h^2}$, with $h$ being a nonzero polynomial in $\R[X]$ and $\mathring{\sigma} \in \mathring{\Sigma}[X]$.\\
We start to recall the famous result by Artin, providing a general solution to Hilbert's 17th problem:
\begin{theorem}{~\cite[Theorem 4]{Artin1927}}
\label{th:hilbertartin}
Let $f \in \R[X]$ be a polynomial non-negative over the reals. Then, $f$ can be decomposed as a sum of squares of rational functions with rational coefficients and there exist a nonzero $h \in \Q[X]$ and $\sigma \in \Sigma[X]$  such that $f = \frac{\sigma}{h^2}$.
\end{theorem}
Given $f \in \R[X]$ non-negative over the reals, let us note $\deg f = d = 2k$,
and $\tau = \tau(f)$. Given $D \in \N$, we denote by $S_D$ the convex hull of the set
\[
\spt{f} + \N^n_{2 D} = \{\alpha + \beta \mid \alpha \in \spt{f}, \beta \in \N^n_{2 D} \} \subseteq \N^n_{d + 2 D}.
\]
Finally, we set $Q_D := S_D / 2 \cap \N^n_{k+D}$.\\
To compute Hilbert-Artin's representation, one can solve the following SDP program:
\begin{align}
\label{eq:dualsdpHA}
\sup_{G, H \succeq 0}  & 
\trace{G}  \\
\text{s.t.} \quad &   \trace{(H \, F_{\gamma})} = \trace{(G \, B_\gamma)}   \,, \quad \forall \gamma \in Q_D  \nonumber \,, \\
\quad & \trace{(H)} = 1  \nonumber \,.
\end{align}
where $B_{\gamma}$ is as for SDP~\eqref{eq:dualsdp}, with rows (resp. columns) indexed by $Q_D$,
and $F_{\gamma}$
has rows (resp.~columns) indexed by $\N_D^n$ with
$(\alpha, \beta)$ entry equal to
$\sum_{\alpha+\beta+\delta = \gamma} f_{\delta}$.
Let us now provide the rationale behind SDP~\eqref{eq:dualsdpHA}. 
The first set of trace equality constraints allows one to find a Gram matrix $H$ associated to $h^2$, with rows (resp.~columns) indexed by $\N_D^n$, as well as a Gram matrix $G$ associated to $\sigma$, with rows (resp.~columns) indexed by $Q_D$.
The last trace equality constraint allows one to ensure that $H$ is not the zero matrix.
Note that we are only interested in finding a stricly feasible solution for SDP~\eqref{eq:dualsdpHA}, thus we can choose any objective function. Here, we maximize the trace, as we would like to obtain a full rank matrix for $G$. 
\begin{proposition}
\label{th:hilbertartinint}
Let $f \in \Z[X]$ be a polynomial non-negative over the reals, with $\deg f = d = 2k$. %and $\tau = \tau(f)$. 
Let us assume that $f$ admits the Hilbert-Artin's representation $f = \frac{\sigma}{h^2}$, with $\sigma \in \mathring{\Sigma}[X]$, $h \in \Q[X]$, $\deg h = D \in \N$ and $\deg \sigma = 2 (D + k)$. Let $Q_D$ be defined as above.
Then, there exist $\mathring{\sigma}_D, \mathring{\sigma} \in \mathring{\Sigma}[X]$ such that
%Then, there exists $N \in \N-\backslash \{ 0 \}$ such that for $\varepsilon := \frac{1}{2^N}$, one has
\[
\mathring{\sigma}_D  f = \mathring{\sigma} \,,
\]
ensuring the existence of a strictly feasible solution $G, H \succ 0$ for SDP~\eqref{eq:dualsdpHA}.
%with $N \leq \tau(\varepsilon) \leq [2 \tau(h) + \tau ] (k+D)^{\bigo{(n)}} $.
\end{proposition}
\begin{proof}
By applying Proposition~\ref{th:boundeps} to $h^2 \, f$, there exists $\varepsilon > 0$  such that $\tilde{\sigma} := h^2 \, f - \varepsilon \sum_{\alpha \in Q_D} X^{2 \alpha}  \in \mathring{\Sigma}[X]$. In addition, for all $\lambda > 0$, one has 
\[
h^2  f = h^2  f + \lambda f \sum_{\alpha \in \N^n_D} X^{2 \alpha} - \lambda f \sum_{\alpha \in \N^n_D} X^{2 \alpha} = \Big( h^2 + \lambda \sum_{\alpha \in \N^n_D} X^{2 \alpha}) \Big) f - \lambda f \sum_{\alpha \in \N^n_D} X^{2 \alpha} = \tilde{\sigma} + \varepsilon \sum_{\alpha \in Q_D} X^{2 \alpha} \,.
\]
Let us define $u_\lambda := \lambda f \sum_{\alpha \in \N^n_D} X^{2 \alpha}$. As in the proof of Proposition~\ref{th:bitmultivsos}, we show that for small enough $\lambda$, the polynomial $\varepsilon \sum_{\alpha \in Q_D} X^{2 \alpha} + u_\lambda$ belongs to ${\Sigma}[X]$. Fix such a $\lambda$, and define $\mathring{\sigma} := \tilde{\sigma} + \varepsilon \sum_{\alpha \in Q_D} X^{2 \alpha} + u_\lambda$ and $\mathring{\sigma}_D := h^2 + \lambda \sum_{\alpha \in \N^n_D} X^{2 \alpha}$. 
Since $\tilde{\sigma} \in \mathring{\Sigma}[X]$, there exists a positive definite Gram matrix $G$  associated to $\mathring{\sigma}$. Similarly, there exists a positive definite Gram matrix $H$  associated to $\mathring{\sigma}_D$. By Theorem~\ref{th:intsospdGram}, this implies that $\mathring{\sigma}, \mathring{\sigma}_D \in \mathring{\Sigma}[X]$, showing the claim. %The second claim is straightforward.
\end{proof}
%
%}
To find such representations in practice, we consider a perturbation of the trace equality constraints of SDP~\eqref{eq:dualsdpHA} where we replace the matrix $G$ by the matrix $G - \epsilon \I$:
\begin{align*}
%\label{eq:dualsdpHAeps}
\mathbf{P}^\varepsilon \quad : \quad \sup_{G, H \succeq 0}  & 
\trace{G}  \\
\text{s.t.} \quad &   \trace{(H \, F_{\gamma})} = \trace{(G  \, B_\gamma)} - \varepsilon \trace{(B_\gamma)}   \,, \quad \forall \gamma \in Q_D  \nonumber \,, \\
\quad & \trace{(H)} = 1  \nonumber \,.
\end{align*}
For $D \in \N$, let us note $\mathring{\Sigma}_D(X) := \{\frac{\sigma}{\sigma_D} : \sigma \in \mathring{\Sigma}[X], \sigma_D \in \Sigma[X] \text{ with } \deg \sigma_D \leq 2 D \}$.\\
Algorithm~$\hilbertsos$ takes as input $f \in \Z[X]$, finds $\sigma_D \in \Sigma[X]$ of smallest degree $2 D$ such that $f \, \sigma_D \in \mathring{\Sigma}[X]$, thanks to an oracle as in~$\intsos$ (i.e., the smallest $D$ for which $f \in \mathring{\Sigma}_D(X)$).
Then, the algorithm finds the largest rational $\varepsilon > 0$ such that Problem~$\mathbf{P}^\varepsilon$ has a strictly feasible solution. 
Problem~$\mathbf{P}^\varepsilon$ is solved by calling the $\sdpHA$ function, relying on an SDP solver. 
Eventually, the algorithm calls the procedure $\absorb$, as in $\intsos$, to recover an exact rational SOS decomposition. }

{
\begin{algorithm}
\caption{{$\hilbertsos$}%: algorithm to compute exact decompositions of positive polynomials into SOS of rational functions.
}
\label{alg:hilbertsos}
{\begin{algorithmic}[1]
%\Require $f \in \mathring{\Sigma}_\Z[X]$, positive $\varepsilon \in \Q$, precision parameters $\delta, R \in \N$ for the SDP solver 
\Require $f \in \Z[X]$ of degree $d = 2 k$, positive $\varepsilon \in \Q$, precision parameters $\delta, R \in \N$ for the SDP solver, precision $\delta_c \in \N$ for the Cholesky's decomposition
lists $\clist_1,\clist_2$ of numbers in $\Q$ and lists $\slist_1,\slist_2$ of polynomials in $\Q[X]$
\State $D \gets 1$
\While {$f  \notin \mathring{\Sigma}[X]/\Sigma_D[X]$} \label{line:DHilberti} 
 $D \gets D + 1$
\EndWhile \label{line:DHilbertf}
\State Compute  the convex hull $S_D$ of $\spt{f} +   \N^n_{d + 2 D}$
\State $Q_D := S_D / 2 \cap \N^n_{k+D}$
\State $t := \sum_{\alpha \in Q_D} X^{2 \alpha}$
\While {Problem~$\mathbf{P}^\varepsilon$ has no strictly feasible solution} \label{line:Hilbertepsi}
 $\varepsilon \gets \frac{\varepsilon}{2}$
\EndWhile \label{line:Hilbertepsf}
\State ok := false
\While {not ok} \label{line:Hilbertdeltai}
\State $(\tilde{G}, \tilde{H}, \tilde{\lambda}_1, \tilde{\lambda}_2) \gets \sdpfunHA{f}{\varepsilon}{ \delta}{R}$ \label{line:Hilbertsdp}
\State $(s_{1 1},\dots,s_{1 r_1}) \gets \choleskyfun{\tilde{G}}{\tilde{\lambda_1}}{\delta_c}$ \label{line:Hilbertchol} 
\State $(s_{2 1},\dots,s_{2 r_2}) \gets \choleskyfun{\tilde{H}}{\tilde{\lambda_2}}{\delta_c}$ 
\State $\tilde{\sigma} := \sum_{i=1}^{r_1} s_{1 i}^2$, $\tilde{\sigma}_D := \sum_{i=1}^{r_2} s_{2 i}^2$
\State $u \gets \tilde{\sigma}_D f - \tilde{\sigma} - \varepsilon t $
\State $\clist_1 \gets [1,\dots,1] $, $\slist_1 \gets [s_{1 1},\dots,s_{r_1 1}]$
\State $\clist_2 \gets [1,\dots,1] $, $\slist_2 \gets [s_{1 2},\dots,s_{r_2 2}]$
\For {$\alpha \in Q_D$}  $\varepsilon_{\alpha} := \varepsilon$
\EndFor
\State $\clist_1, \slist_1,(\varepsilon_\alpha) \gets \absorbfun{u}{Q_D}{(\varepsilon_\alpha)}{\clist_1}{\slist_1}$ \label{line:Hilbertabsorb}
\If {$\min_{\alpha \in Q_D} \{ \varepsilon_\alpha \} \geq 0$} ok := true \label{line:Hilbertsosok}
\Else $\ \delta \gets 2 \delta$, $R \gets 2 R$, $\delta_c \gets 2 \delta_c$
\EndIf
\EndWhile \label{line:Hilbertdeltaf}
\For {$\alpha \in Q_D$}   $\clist_1 \gets  \clist_1 \cup \{ \varepsilon_\alpha \}$, $\slist_1 \gets  \slist_1 \cup \{ X^\alpha \}$
\EndFor
\State \Return $\clist_1$, $\clist_2$, $\slist_1$, $\slist_2$
\end{algorithmic}}
\end{algorithm}
}

{
\begin{theorem}
\label{th:hilbertsos}
Let $f \in \Z[X] \cap \mathring{\Sigma}_D(X)$ and assume that the SOS polynomials involved in the denominator of $f$ have coefficients of bit size at most $\tau_D \geq \tau$. 
On input $f$, Algorithm~$\hilbertsos$ terminates and outputs a weighted SOS
decomposition for $f$. 
There exist $\varepsilon, \delta, R, \delta_c$ of bit sizes upper bounded by  \revise{$\tau_D \cdot (d+D)^{(d+D)^{\bigo{( n)}  }} $} such that $\hilbertsosfun{f}{\varepsilon}{\delta}{R}{\delta_c}$ runs in boolean running time \revise{$\tau_D^2 \cdot (d+D)^{(d+D)^{\bigo{( n)}  }}  $}.
%
%The maximum bit size of its coefficients
%involved in this SOS decomposition is upper bounded by $\tau_D D^{\bigo{(n)}}$ and the boolean running time of the
%procedure is upper bounded by $\tau_D^2 D^{\bigo{(n)}}$.
\end{theorem}
\begin{proof}
Since $f \in \mathring{\Sigma}_D(X)$, the first loop of Algorithm~$\hilbertsos$ terminates and there exists a strictly feasible solution for SDP~\eqref{eq:dualsdpHA}, by Proposition~\eqref{th:hilbertartinint}. Thus, there exists a small enough $\varepsilon > 0$ such that Problem~$\mathbf{P}^\varepsilon$ has also a strictly feasible solution. This ensures that the second loop of Algorithm~$\hilbertsos$ terminates. Then, one shows as for Algorithm~$\intsos$ that the absorption procedure succeeds, yielding termination of the third loop.
Let us note \[\Delta_D(u) := \{(\alpha, \beta) : \alpha + \beta \in \spt{u} \,,
\alpha,\beta \in Q_D\,, \alpha \neq \beta \} \,. \]
The first output $\clist_1$ of Algorithm~$\hilbertsos$ is a 
list of non-negative rational numbers containing the list $[1, \dots, 1]$ of length $r_1$, the list
$\bigl\{\frac{|u_{\alpha + \beta}|}{2} : (\alpha, \beta) \in \Delta_D(u)
\bigr\}$
and the list $\{ \varepsilon_\alpha : \alpha \in Q_D \}$.  The
second output $\slist_1$ of Algorithm~$\intsos$ is a list of polynomials 
containing the list $[s_{1 1}, \dots, s_{r_1 1}]$, the list
$\{ X^\alpha + \sgn{(u_{\alpha+\beta})} X^\beta : (\alpha, \beta) \in
\Delta_D(u) \}$
and the list $\{ X^\alpha : \alpha \in Q_D \}$. From these two outputs, one reconstructs the weighted SOS decomposition of the numerator $\sigma$ of $f$.
The third output $\clist_2$ is a list of non-negative rational  numbers containing the list $[1, \dots, 1]$ of length $r_2$ and the fourth output is a list of polynomials $[s_{1 2}, \dots, s_{r_2 2}]$. From these two outputs, one reconstructs the weighted SOS decomposition of the denominator $\sigma_D$ of $f$.
At the end, we obtain the weighted SOS decomposition $f = \frac{\sigma}{\sigma_D}$ with
\[
\sigma_D := \sum_{i=1}^{r_2} s_{2 i}^2 \,, \quad 
\sigma :=  \sum_{i=1}^{r_1} s_{1 i}^2 + \sum_{\mbox{\tiny
    $(\alpha, \beta) \in \Delta_D(u)$ }} \dfrac{|u_{\alpha + \beta}|}{2}
(X^\alpha + \sgn{(u_{\alpha+\beta})} X^\beta)^2 + \sum_{\mbox{\tiny
    $\alpha \in Q_D$ }} \varepsilon_\alpha X^{2 \alpha}  \,. \]
%    
%Writing $f = \frac{\sigma}{\sigma_D}$, one shows as in Proposition~\ref{th:boundeps} that the largest rational number belonging to the set $\{ \varepsilon \in \R^{> 0} : \forall \bmx \in \R^n, \sigma_D(\bmx) f(\bmx)  - \varepsilon \sum_{\alpha \in Q_D} \bmx^{2 \alpha}\geq 0 \}$ has bit size upper bounded by $\bigo{( \tau_D \cdot (4 d + 4 D +2)^{3n+3} )}$. 
We conclude our bit complexity analysis as in Proposition~\ref{th:bitmultivsos} and Theorem~\ref{th:costmultivsos}.
\end{proof}
\begin{remark}
\label{rk:HAbounds}
Note that even if the bit complexity of $\hilbertsos$ is singly exponential in the degree $D$ of the denominator, this degree can be rather large. In~\cite{Lombardi18}, the authors provide an upper bound expressed with a tower of five exponentials for  the degrees of denominators involved in Hilbert-Artin's representations.
\end{remark}
}
%
%%% Local Variables:
%%% mode: latex
%%% TeX-master: "multivsos"
%%% End:
\section{Exact Putinar's representations}
\label{sec:putinar}
We let $f, g_1,\dots, g_m$ in $\Z[X]$ of degrees less than $d \in \N$ and $\tau \in \N$ be
a bound on the bit size of their coefficients. Assume that $f$ is
positive over
$\K := \{\x \in \R^n : g_1(\x) \geq 0, \dots, g_m(\x) \geq 0 \}$ and
reaches its infimum with $f^\star := \min_{\x \in \K} f(\x) > 0$.  With
$f = \sum_{|\alpha| \leq d} f_\alpha \x^\alpha$, we set
$\|f\| := \max_{|\alpha| \leq d} \frac{f_\alpha \alpha_1!\cdots
  \alpha_n!}{|\alpha|!}$ and $g_0 := 1$. 
  
  We consider the quadratic module
$\mathcal{Q}(\K) := \bigl\{ \sum_{j=0}^m \sigma_j g_j : \sigma_j \in
\Sigma[\x] \bigl\}$
and, for $D\in \N$, the $D$-truncated quadratic module
$\mathcal{Q}_D(\K) := \bigl\{ \sum_{j=0}^m \sigma_j g_j : \sigma_j \in
\Sigma[\x] \,, \ \deg(\sigma_j g_j) \leq D \bigl\}$
generated by $g_1,\dots, g_m$.  We say that $\mathcal{Q}(\K)$ is {\em
  archimedean} if $N - G_n \in \mathcal{Q}(\K)$ for some $N \in
\N$ (recall that $G_n := \sum_{i=1}^n X_i^2$). 
We also assume in this section:
\begin{assumption}
\label{hyp:arch}
The set $\K$ is a basic compact semi-algebraic set with nonempty
interior, included in $[-1, 1]^n$
and %involving the constraints $1 - x_1^2 \geq 0, \dots, 1 - x_n^2 \geq 0$, ensuring
$\mathcal{Q}(\K)$ is archimedean.
\end{assumption}
%

%If $f\in\mathcal{Q}_D(\K)$ for some $D \in 2 \N$, then one can write
%$f = \sum_{i=0}^m \sigma_j g_j$ with $\sigma_j \in \Sigma[X]$ for 
%$0\leq j\leq m$, thus $f$ is positive over $\K$. The converse
%implication is due to Putinar.

Under Assumption~\ref{hyp:arch}, $f$ is positive over $\K$ only if
$f\in\mathcal{Q}_D(\K)$ for some $D \in 2 \N$
(see~\cite{Putinar1993positive}). In this case, there exists a {\em
  Putinar's representation} $f = \sum_{i=0}^m \sigma_j g_j$ with
$\sigma_j \in \Sigma[X]$ for $0\leq j\leq m$.
% MS
% For instance, with
% $f =-X_1^2 - 2 X_1 X_2 - 2 X_2^2 + 6$ and
% $S := \{(\x_1,\x_2) \in \R^2 : 1 - \x_1^2 \geq 0, 1 - \x_1^2 \geq
% 0\}$,
% we obtain the Putinar's representation
% $f = (1 + X_1-X_2)^2 + 2 (1-X_1^2) + 3 (1 - X_2^2)$.
% %
%
%Let $w_j := \lceil \deg g_j / 2 \rceil$, for all $1\leq j\leq m$.  
%
% In the unconstrained case, one can solve
% SDP~\eqref{eq:dualsdp} to certify that $f \in \Sigma[\x]$. In the
% constrained case, o
One can certify that $f \in \mathcal{Q}_D(\K)$ for $D = 2k$ by solving
the next SDP with
$k \geq \lceil d / 2 \rceil$:
\if{
\begin{equation}
\label{eq:primalsdp2}
\inf\limits_{\y} \quad  \sum_{|\alpha| \leq D} f_\alpha y_{\alpha}  \quad \text{s.t.}\quad 
 \begin{aligned}
 M_k(\y) \succeq 0 \,, \quad & \quad y_0 = 1  \,, \\
 M_{k-w_j}(g_j\, \y) \succeq 0 \,, & \quad j=1,\dots,m  \,. \\
\end{aligned}
\end{equation}
By writing $M_k(\y) = \sum_{\alpha} y_\alpha B_\alpha$ and
$M_{k-w_j}(g_j \, \y) = \sum_{\alpha } y_\alpha C_{j \alpha}$ for all
$j=1,\dots,m$ , the dual of SDP~\eqref{eq:primalsdp2} is also an SDP
given by:
}\fi
\begin{align}
\label{eq:dualsdp2}
\inf_{G_0, G_1,\dots,G_m \succeq 0}  & 
\trace{(G_0 \, B_0)} + \sum_{i=1}^m g_j(0) \trace{(G_j \, C_{j0})} \\
\text{s.t.} \quad &  \trace{(G_0 \, B_\gamma)} + \sum_{j=1}^m \trace{(G_j \, C_{j \gamma})} = f_\gamma \,, \quad \forall \gamma \in \N_D^n - \{0\} \nonumber \,,
\end{align}
where $B_{\gamma}$ is as for SDP~\eqref{eq:dualsdp} and $C_{j \gamma}$
has rows (resp.~ columns) indexed by $\N_{k-w_j}^{n}$ with
$(\alpha, \beta)$ entry equal to
$\sum_{\alpha+\beta+\delta = \gamma} g_{j \delta}$.
SDP~\eqref{eq:dualsdp2} is a reformulation of the problem
\[ f_D^\star := \sup \{b : f - b \in \mathcal{Q}_D(\K) \} \,. \]
Thus $f_D^\star$ is also the optimal value of SDP~\eqref{eq:dualsdp2}.  
% As for Theorem~\ref{th:lasunconstrained}, we
% have the next result following directly
% from~\cite[Theorem~4.2]{Las01sos}.
The next result follows from \cite[Theorem~4.2]{Las01sos}:
\begin{theorem}
\label{th:lasoncstrained}
We use the notation and assumptions introduced above.  For
$D \in 2\N$ large enough, one has 
\[
0 < f_D^\star \leq f^\star 
\,. \]
In addition, SDP~\eqref{eq:dualsdp2} has an optimal solution
$(G_0,G_1,\dots,G_m)$, yielding the following Putinar's
representation:
\[
f - f_D^\star = \sum_{i=1}^r \lambda_{i0} q_{i0}^2 + \sum_{i=1}^m g_j
\sum_{i=1}^{r_j} \lambda_{i j} q_{i j}^2 \,,
\]
where the vectors of coefficients of the polynomials $q_{ij}$ are the
eigenvectors of $G_j$ with respective eigenvalues $\lambda_{ij}$, for
all $j=0,\dots,m$.
\end{theorem}
The complexity of Putinar's Positivstellens\"atz was analyzed {by~\cite{Nie07Putinar}}:
\begin{theorem}
\label{th:putinar}
With the notation and assumptions introduced above,   there
exists a real $\chi_\K > 0$ depending on $\K$ such that

(i) for all even
$D \geq \chi_\K \exp \bigl(d^2 n^d \frac{\|f\|}{f^\star} \bigr)^{\chi_\K}$,
$f \in \mathcal{Q}_D(\K)$.

(ii) for all even $D \geq \chi_\K \exp \bigl(2 d^2 n^d \bigr)^{\chi_\K}$,
$0 \leq f^\star - f_D^\star \leq \frac{6 d^3 n^{2 d}
  \|f\|}{\sqrt[\chi_\K]{\log \frac{D}{\chi_\K}}}$.
%$0 \leq f^\star - f_D^\star \leq \frac{6 d^3 n^{2 d} \|f\|}{\sqrt[c]{\log \frac{D}{c}}}$.
\end{theorem}
From a computational viewpoint, one can certify that $f$ lies in $\mathcal{Q}_D(\K)$ for
$D = 2k$ large enough, by solving SDP~\eqref{eq:dualsdp2}. Next, we show how to ensure the existence of a strictly feasible solution for
SDP~\eqref{eq:dualsdp2} after replacing the set defined by our initial constraints
$\K$ by the following one
\[
\K':= \{ \bmx \in \K : 1 - \bmx^{2 \alpha} \geq 0 \,, \forall \alpha \in
\N^n_k \}
\,. \]
\subsection{Preliminary results}
\label{sec:prelimputinar}
We first give a lower bound for $f^\star$.
\begin{proposition}
\label{th:lowerboundcube}
\revise{With the above notation and assumptions, one has
\[f^\star \geq 2^{-\tau m^{n+1} d^{O(n)}}
   \,.
  \]}
\end{proposition}
\begin{proof}
\revise{
By \cite[\S~14.1-14.2]{BPR06}, one obtains a univariate polynomial of degree $m^{n+1} d^{O(n)}$ of bit size $\tau d^{O(n)}$ for which $f^\star$ is a nonzero root. 
The desired result then follows by Lemma \ref{th:cauchyint}. 
}  
\end{proof}
\begin{theorem}
\label{th:putinarint}
We use the notation and assumptions introduced above.  There exists
$D \in 2 \N$ such
that:\\
(i) $f \in \mathcal{Q}_D(\K)$ with the representation
\[f = f_D^\star + \sum_{j=0}^m \sigma_j g_j \,, \]
 for $f_D^\star > 0$,
$\sigma_j \in \Sigma[X]$ with $\deg (\sigma_j g_j) \leq D$ for all
$j=0,\dots,m$.\\
(ii) $f \in \mathcal{Q}_D(\K')$ with the representation
\[f = \sum_{j=0}^m \mathring{\sigma_j} g_j + \sum_{|\alpha| \leq k}
c_\alpha (1 - X^{2 \alpha}) 
\,,\]
for $\mathring{\sigma_j} \in \mathring{\Sigma}[X]$ with
$\deg (\mathring{\sigma_j} g_j) \leq D$, for all $j=0,\dots,m$, and
some sequence of positive numbers $(c_\alpha)_{|\alpha| \leq k}$.\\
(iii) {There exists a real $C_S > 0$ depending on $S$ and
$\varepsilon = \frac{1}{2^N}$ with positive $N \in \N$ such that
\[f - \varepsilon \sum_{|\alpha| \leq k} X^{2 \alpha} \in
\mathcal{Q}_D(\K') \,, \quad \revise{N \leq 2^{\tau d^{C_S n }}} \,,\] 
}where $\tau$ is the maximal bit size
of the coefficients of $f,g_1,\dots,g_m$.
\end{theorem}
\begin{proof}
  Let $\chi_\K$ be as in Theorem~\ref{th:putinar} and $D = 2 k$ be the
  smallest integer larger than
  \[
  \underline{D} := \max\{\chi_\K \exp \Bigl(\frac{12 d^3 n^{2d}
    \|f\|}{f^\star} \Bigr)^{\chi_\K}, \chi_\K ( 2 d^2 n^d )^{\chi_\K} \} \,. \]
  Theorem~\ref{th:putinar} implies that $f \in \mathcal{Q}_D(\K)$
  and
  $f^\star - f_{D}^\star \leq \frac{6 d^3 n^{2 d}
    \|f\|}{\sqrt[\chi_\K]{\log \frac{{D}}{\chi_\K}}} \leq
  \frac{f^\star}{2}$.

  (i) This yields the representation
  $f - f_D^\star = \sum_{j=0}^m \sigma_j g_j$, with
  $f_D^\star \geq \frac{f^\star}{2} > 0$, $\sigma_j \in \Sigma[X]$ and
  $\deg (\sigma_j g_j) \leq D$ for all $j=0,\dots,m$.

  (ii) For $1\leq j\leq m$, let us define
  \[ t_j := \sum_{|\alpha| \leq k - w_j} X^{2 \alpha} \,, \quad
  t_0 := \sum_{|\alpha| \leq k} X^{2 \alpha} \,, \quad 
  t := \sum_{j=0}^m t_j g_j \,. \]
  For a given $\nu > 0$, we use the
  perturbation polynomial
  $-\nu t = -\nu \sum_{|\gamma|\leq D}t_\gamma X^\gamma$. 
  For each term $-t_\gamma X^\gamma$, one has $\gamma = \alpha + \beta$ with
  $\alpha, \beta \in \N^n_k$, thus
  \[
  -t_\gamma X^\gamma = |t_\gamma|(-1 + \frac{1}{2} (1 - X^{2
    \alpha}) + \frac{1}{2} (1 - X^{2 \beta}) + \frac{1}{2} (X^\alpha -
  \sgn{(t_\gamma)} X^\beta)^2) \,. \]
As in the proof of Proposition~\ref{th:bitmultivsos}, let us note
\[
\Delta(t) := \{(\alpha, \beta) : \alpha + \beta \in \spt{t} \,,
  \alpha,\beta \in \N^n_k \,, \alpha \neq \beta \}
 \,. \]
  Hence, for all $\alpha \in \N^n_k$, there exists $d_\alpha \geq 0$  such
  that 
  \[f = f - \nu t + \nu t =
   f_D^\star - \sum_{|\gamma|\leq D} \nu |t_\gamma| + \sum_{j=0}^m
  \sigma_j g_j + \nu t + \sum_{{ |\alpha| \leq k }} d_\alpha (1 -
  X^{2 \alpha}) + \nu \sum_{ (\alpha, \beta) \in \Delta(t) }
  \dfrac{|t_{\alpha + \beta}|}{2} (X^\alpha -
  \sgn{(t_{\alpha+\beta})} X^\beta)^2  \,. \]
Since one has not necessarily $d_\alpha > 0$ for all $\alpha \in \N^n_k$, we now explain how to 
  handle the case when $d_\alpha = 0$ for 
  $\alpha \in \N^n_k$. We write
\begin{align*}
  - \sum_{|\gamma|\leq D} \nu |t_\gamma| + \sum_{{ |\alpha| \leq k }}
  d_\alpha (1 - X^{2 \alpha})= & 
  - \sum_{|\gamma|\leq D} \nu |t_\gamma|
  - \sum_{\alpha : d_\alpha = 0} \nu + \sum_{\alpha : d_\alpha = 0}
  \nu (1 - X^{2\alpha}) 
  + \sum_{\alpha : d_\alpha = 0} \nu X^{2
    \alpha} \\
    & + \sum_{{ |\alpha| : d_\alpha = 0 }} d_\alpha (1 - X^{2
    \alpha}) + \sum_{{ |\alpha| : d_\alpha > 0 }} d_\alpha (1 - X^{2
    \alpha}) \,. 
\end{align*}
For $\alpha \in \N^n_k$, we define $c_\alpha := \nu$ if
  $d_\alpha = 0$ and $c_\alpha := d_\alpha$ otherwise, 
  $a := \sum_{|\gamma|\leq D} |t_\gamma| + \sum_{\alpha : d_\alpha =
    0} 1$,
  $\mathring{\sigma}_j := \sigma_j + \nu t_j$, for each $j=1,\dots,m$
  and
  \[\mathring{\sigma}_0 := f_D^\star - \nu a + \sigma_0 + \nu t_0 + \nu
  \sum_{ (\alpha, \beta) \in \Delta(t) } \dfrac{|t_{\alpha +
      \beta}|}{2} (X^\alpha - \sgn{(t_{\alpha+\beta})} X^\beta)^2 +
  \sum_{\alpha : d_\alpha = 0} \nu X^{2 \alpha} \,. \]
    So, there
  exists a sequence of positive numbers $(c_\alpha)_{|\alpha|\leq k}$
  such that
  \[
  f = \sum_{j=0}^m \mathring{\sigma_j} g_j + \sum_{{ |\alpha| \leq k
    }} c_\alpha (1 - X^{2 \alpha}) \,.
    \]
%
%
%By writing $\sum_{j=0}^m \sigma_j g_j + \nu t = \sum_{j=0}^m (\sigma_j + \nu t_j) g_j$. 
  Now, let us select $\nu := \frac{1}{2^M}$ with $M$ being the
  smallest positive integer such that
  $0 < \nu \leq \frac{f_D^\star}{2 a}$. 
 This implies the existence of a positive definite Gram matrix for $\mathring{\sigma_0}$, thus by Theorem~\ref{th:intsospdGram}, 
  $\mathring{\sigma_0} \in \mathring{\Sigma}[X]$. Similarly, for
  $1\leq j\leq m$, $\mathring{\sigma}_j$  belongs to
  $\mathring{\Sigma}[X]$, which proves the second claim.

  (iii) Let $N := M+1$ and
  $\varepsilon := \frac{1}{2^N} = \frac{\nu}{2}$. One has
  \[f - \varepsilon \sum_{|\alpha| \leq k} X^{2 \alpha} = f -
  \varepsilon t_0 = \mathring{\sigma}_0 - \varepsilon t_0 +
  \sum_{j=1}^m \mathring{\sigma}_j g_j + \sum_{|\alpha| \leq k}
  c_\alpha (1 - X^{2 \alpha}) \,. \]
  Thus, $\sigma_0 + (\nu - \varepsilon) t_0 \in
  \mathring{\Sigma}[X]$.
  This implies that
  $\mathring{\sigma}_0 - \varepsilon t_0 \in \mathring{\Sigma}[X]$ and
  $f - \varepsilon t_0 \in \mathcal{Q}_D(\K')$. Next, we derive a
  lower bound of $\frac{f^\star_D}{a}$.
  Since
  \[t =\sum_{|\alpha|\leq k} X^{2\alpha} + \sum_{j=1}^m g_j
  \sum_{|\alpha| \leq k - w_j} X^{2\alpha} \,,\]
  one has
  \[\sum_{|\gamma| \leq D} |t_\gamma| \leq 2^\tau (m+1)
  \binom{n+D}{n} \,. \]
  This implies that
\[ a \leq 2^{\tau} (m+1) \binom{n+D}{n} + \binom{n+k}{k} \leq 2^{\tau}
  (m+2) \binom{n+D}{n} \,. \]
  Recall that $\frac{f^\star}{2} \leq f^\star_D$, {implying
  \[\frac{f^\star_{{D}}}{a} \geq \frac{f^\star }{2^{\tau+1} (m+2)
    \binom{n+{D}}{n}} \geq \frac{1}{(m+2) 2^{- \revise{\tau m^{n+1} d^{O(n})} }
    D^n} \,, \]
}
  where the last inequality follows from
  Proposition~\ref{th:lowerboundcube}.  Let us now give an upper bound of
  $\log_2 D$. First, note that for all $\alpha \in \N^n$,
  $\frac{|\alpha|!}{\alpha_1!\cdots\alpha_n!} \geq 1$, thus
  $\|f\| \leq 2^\tau$.
  Since $D$ is the smallest even integer larger than $\underline{D}$,
  one has {
  \[\log_2 D \leq 1 + \log_2 \underline{D} \leq 1 + \log \chi_\K + (12 d^3
  n^{2 d} 2^\tau 2^\revise{{\tau m^{n+1} d^{O(n)}}}  )^{\chi_\K} \,.\]
  Next, since $N$ is the smallest integer such that
  $\varepsilon = \frac{1}{2^N} = \frac{\nu}{2} \leq \frac{f^\star_D}{2
    a}$,
  it is enough to take
  \[ \revise{N \leq
  2^{\tau d^{C_S  n }}} \,,\]
  for some real $C_S > 0$ depending on $S$, the desired result.}
\end{proof}
\subsection{Algorithm~$\putinarsos$}
\label{sec:putinarsos}

We can now present Algorithm~$\putinarsos$. \\

\begin{algorithm}
  \caption{$\putinarsos$.% : algorithm to compute exact rational
    % Putinar's representations of polynomials positive over basic
    % compact semi-algebraic sets
    }
\label{alg:putinarsos}
\begin{algorithmic}[1]
\Require $f,g_1,\dots,g_m \in \Z[X]$ of degrees less than $d \in \N$, $\K := \{\x \in \R^n : g_1(\x) \geq 0, \dots, g_m(\x) \geq 0 \}$, positive $\varepsilon \in \Q$, precision parameters $\delta, R \in \N$ for the SDP solver 
,  precision $\delta_c \in \N$ for the Cholesky's decomposition
\Ensure lists $\clist_0,\dots,\clist_m, \calpha$ of numbers in $\Q$ and lists $\slist_0,\dots,\slist_m$ of polynomials in $\Q[X]$
\State $k \gets  \lceil d / 2 \rceil$, $D \gets 2 k$, $g_0 := 1$
\While {$f  \notin \mathcal{Q}_D(\K)$} \label{line:DPuti} 
 $k \gets k + 1$, $D \gets D + 2$
\EndWhile \label{line:DPutf}
\State $P := \N^n_D$, $Q := \N^n_k$,  $\K':= \{ \x \in \K : 1 - \x^{2 \alpha} \geq 0 \,, \forall \alpha \in \N^n_k \}$ \label{line:npP}
\State $t := \sum_{\alpha \in Q} X^{2 \alpha}$, $f_\varepsilon \gets f - \varepsilon t$
\While {$f_\varepsilon \notin  \mathcal{Q}_D(\K')$} \label{line:epsiP}
 $\varepsilon \gets \frac{\varepsilon}{2}$, $f_\varepsilon \gets f - \varepsilon t$
\EndWhile \label{line:epsfP}
\State ok := false
\While {not ok} \label{line:deltaiP}
\State $[\tilde{G_0},\dots,\tilde{G}_{m},  \tilde{\lambda}_0,\dots,\tilde{\lambda}_{m},(\tilde{c}_\alpha)_{|\alpha| \leq k}], \gets \sdpconfun{f_\varepsilon}{ \delta}{R}{\K'}$ \label{line:sdpcon}
\State $\calpha \gets (\tilde{c}_\alpha)_{|\alpha| \leq k}$
\For {$j \in \{0,\dots,m\}$}
\State $(s_{1j},\dots,s_{r_j j}) \gets \choleskyfun{\tilde{G}_j}{\tilde{\lambda}_j}{\delta_c}$, $\tilde{\sigma}_j := \sum_{i=1}^{r_j} s_{i j}^2 $ \label{line:cholcon}
\State $\clist_j \gets [1,\dots,1] $, $\slist_j \gets [s_{1j},\dots,s_{r_j j}]$
\EndFor
\State $u \gets f_\varepsilon - \sum_{j=0}^m   \tilde{\sigma_j} \, g_j - \sum_{|\alpha| \leq k} \tilde{c}_\alpha (1 - X^{2 \alpha})$
\For {$\alpha \in Q$}  $\varepsilon_{\alpha} := \varepsilon$
\EndFor
\State $\clist, \slist,(\varepsilon_\alpha) \gets \absorbfun{u}{Q}{(\varepsilon_\alpha)}{\clist}{\slist}$ \label{line:absorbP}
\If {$\min_{\alpha \in Q} \{ \varepsilon_\alpha \} \geq 0$} ok := true 
\Else $\ \delta \gets 2 \delta$, $R \gets 2 R$, $\delta_c \gets 2 \delta_c$
\EndIf
\EndWhile \label{line:deltafP}
\For {$\alpha \in Q$}  
\State $\clist_0 \gets  \clist_0 \cup \{ \varepsilon_\alpha \}$, $\slist_0 \gets  \slist_0 \cup \{ \x^\alpha \}$
\EndFor
\State \Return $\clist_0,\dots,\clist_m, \calpha,  \slist_0,\dots,\slist_m$
\end{algorithmic}
\end{algorithm}

For $f \in \Z[X]$ positive
over a basic compact semi-algebraic set $S$ satisfying
Assumption~\ref{hyp:arch}, the first loop % from line~\lineref{line:Di}
% to line~\lineref{line:Df}
outputs the smallest positive integer
$D = 2k$ such that $f \in \mathcal{Q}_D(\K)$. \\
Then the procedure is
similar to $\intsos$. As for the first loop of
$\intsos$, the loop from line~\lineref{line:epsiP} to
line~\lineref{line:epsfP} allows us to obtain a perturbed polynomial
$f_\varepsilon \in \mathcal{Q}_D(\K')$, with
$\K' := \{ \bmx \in \K : 1 - \bmx^{2 \alpha} \geq 0 \,, \forall \alpha
\in \N^n_k \}$.\\
Then one solves SDP~\eqref{eq:dualsdp2} with the $\sdpcon$ procedure
and performs Cholesky's decomposition to obtain an approximate
Putinar's representation of $f_\varepsilon = f - \varepsilon t$
and a remainder $u$. \\
Next, we apply the $\absorb$ subroutine as 
in $\intsos$. The rationale is that with large
enough precision parameters for the procedures $\sdpcon$ and
$\cholesky$, one finds an exact weighted SOS decomposition of
$u + \varepsilon t$, which yields in turn an exact Putinar's
representation of $f$ in $\mathcal{Q}_D(\K')$ with rational
coefficients.
\begin{example}
\label{ex:putinar}
Let us apply $\putinarsos$ to $f =-X_1^2 - 2 X_1 X_2 - 2 X_2^2 + 6$,
$S := \{(x_1,x_2) \in \R^2 : 1 - x_1^2 \geq 0, 1 - x_2^2 \geq 0\}$
and the same precision parameters as in Example~\ref{ex:intsos}. The
first and second loop yield $D = 2$ and 
% and $S' = S$. 
$\varepsilon = 1$. After running $\absorb$, we obtain the exact
Putinar's representation 
\[f = \frac{23853407}{292204836} + \frac{23}{49} X_1^2 +
\frac{130657269}{291009481} X_2^2 + \frac{1}{2442^2} + (X_1-X_2)^2 +
(\frac{X_2}{2437})^2+(\frac{11}{7})^2 (1-X_1^2) + (\frac{13}{7})^2
(1-X_2^2) \,.\]
\end{example}
\subsection{Bit complexity analysis}
\label{sec:bitputinar}
\begin{theorem}
\label{th:putinarsos}
We use the notation and assumptions introduced above.  
For some constant $C_\K > 0$ depending on $S$, there exist $\varepsilon$ and $D = 2 k$ of bit sizes less than \revise{${2^{\tau d^{C_S n}} }$}, and 
$\delta$, $R$, $\delta_c$ of bit sizes less than 
$((m+1) D^n)^{\bigo{(D^n)}}$ for which
$\putinarsosfun{f}{\K}{\varepsilon}{\delta}{R}{\delta_c}$ terminates
and outputs an exact Putinar's representation with rational
coefficients of $f \in \mathcal{Q}(\K')$, with
$\K' := \{ \bmx \in \K : 1 - \bmx^{2 \alpha} \geq 0 \,, \forall \alpha
\in \N^n_k \}$.
\revise{
The maximum bit size of these coefficients and the boolean running time are both upper bounded by $((m+1) D^n)^{\bigo{(D^n)}}$.
}
\if{
The maximum bit size of these coefficients is bounded by
$\bigo{(  2^{C_S \tau d^{2 n + 2}}   )}$ and the procedure runs in boolean time
$\bigo{\bigl(2^{   2^{K_S \tau d^{2 n + 2}}   } \bigr)}$.}
}\fi
\end{theorem}
\begin{proof}
{
  The loops going from line~\lineref{line:DPuti} to
  line~\lineref{line:DPutf} and from line~\lineref{line:epsiP} to
  line~\lineref{line:epsfP} always terminate as respective
  consequences of Theorem~\ref{th:putinarint}~(i) and
  Theorem~\ref{th:putinarint}~(iii) with $\log_2 D \leq \revise{{2^{\tau d^{C_S n}} }} $,
  $\varepsilon = \frac{1}{2^N}$, $N \leq \revise{{2^{\tau d^{C_S n}} }} $, for
  some real $C_\K > 0$ depending on $\K$.
  What remains to prove is similar to
  Proposition~\ref{th:bitmultivsos} and
  Theorem~\ref{th:costmultivsos}.  \\
  Let $\nu$,
  $\mathring{\sigma}_0,\dots,\mathring{\sigma}_m, (c_\alpha)_{|\alpha|
    \leq k}$
  be as in the proof of Theorem~\ref{th:putinarint}. Note that $\nu$
  (resp.~$\varepsilon-\nu$) is a lower bound of the smallest
  eigenvalues of any Gram matrix associated to $\mathring{\sigma}_j$
  (resp.~$\mathring{\sigma}_0$) for $1\leq j\leq m$. In addition,
  $c_\alpha \geq \nu$ for all $\alpha \in \N^n_k$.  When the $\sdp$
  procedure at line~\lineref{line:sdpcon} succeeds, the matrix
  $\tilde{G}_j$ is an approximate Gram matrix of the polynomial
  $\mathring{\sigma}_j$ with $\tilde{G}_j \succeq 2^{\delta} I $,
  $\sqrt{\trace{(\tilde{G}_j^2)}} \leq R$, we obtain a positive
  rational approximation $\tilde{\lambda}_j \geq 2^{-\delta}$ of the
  smallest eigenvalue of $\tilde{G}_j$, $\tilde{c_\alpha}$ is a
  rational approximation of $c_\alpha$ with
  $\tilde{c_\alpha} \geq 2^{-\delta}$, and $\tilde{c_\alpha} \leq R$,
  for all $j=0,\dots,m$ and $\alpha \in \N^n_k$.  This happens when
  $2^{-\delta} \leq \varepsilon$ and
  $2^{-\delta} \leq \varepsilon - \nu$, thus for
  $\delta = \bigo{(  \revise{{2^{\tau d^{C_S n}} }}  )  }$. \\
\revise{
Next, we will give an upper bound on the bit size of $R$.
Note that the size $r_0$ of the matrix $\tilde{G_0}$ satisfies
  $r_0 \geq r_j$ for all $j=1,\dots,m$. 
  In addition, $\deg g_j \leq d$
  implies
  \[\sum_\delta |g_{j \delta}| \leq \binom{n+\deg g_j}{n} 2^\tau \leq
  \binom{n+d}{n} 2^\tau \leq d^n 2^{\tau+1} \,. \]
Thus, the bit size of the entries of each matrix $C_{j, \gamma}$ is upper bounded by $\lceil n \log_2 d \rceil + \tau + 1 = \bigo{(D)}$.
As in the proof of Theorem~\ref{th:costmultivsos}, let $\nsdp$ be
  the sum of the sizes of the matrices involved in
  SDP~\eqref{eq:dualsdp2} and $\msdp$ the number of equality constraints. Note
  that 
  \[\nsdp \leq (m+1) r_0 \leq (m+1) \binom{n+D}{n} \,, \quad \msdp := \binom{n+D}{n} \,. \]  
Therefore, as in the proof of Proposition~\ref{th:boundR}, we obtain an upper bound of $((m+1) D^n)^{\bigo{(D^n)}} = ((m+1) D)^{D^{\bigo{(n)}} } $ for the bit size of $R$.
}
\if{
  As in the proof of
  Proposition~\ref{th:boundR}, we derive a similar upper bound of $R$
  by a symmetric argument while considering a Putinar representation
  of $\overline{f}_D - f \in \mathcal{Q}_D(\K')$, where
  \[\overline{f}_D := \inf \{b : b - f \in \mathcal{Q}_D(\K) \} \,. \]  
  }\fi
  As for the second loop of Algorithm~$\intsos$, the third loop of
  $\putinarsos$ terminates when the remainder polynomial
  \[
  u = f_\varepsilon - \sum_{j=0}^m \tilde{\sigma_j} \, g_j -
  \sum_{|\alpha| \leq k} \tilde{c}_\alpha (1 - X^{2 \alpha}) \]
  satisfies $|u_\gamma| \leq \frac{\varepsilon}{r_0}$, where
  $r_0 = \binom{n+k}{n}$ is the size of $Q = \N^n_k$. As in the
  proof of Proposition~\ref{th:bitmultivsos}, one can show that this
  happens when $\delta$ and $\delta_c$ are large enough.
  To bound the precision $\delta_c$ required for Cholesky's
  decomposition, we do as in the proof of
  Proposition~\ref{th:bitmultivsos}.
%Note that one has $\trace{(\tilde{G}_0 B_\gamma)} = \sum_{\alpha + \beta = \gamma} \tilde{G}_{0 \alpha,\beta}$  and $\trace{(\tilde{G}_j C_{j\gamma})} = \sum_\delta g_\delta \sum_{\alpha + \beta + \delta = \gamma} \tilde{G}_{j \alpha,\beta}$.
%
  The difference now is that there are $m + \binom{n+k}{k} = m + r_0$
  additional terms in each equality constraint of
  SDP~\eqref{eq:dualsdp2}, by comparison with SDP~\eqref{eq:dualsdp}.
  Thus, we need to bound for all $j=1\,\dots,m$, $\alpha \in \N^n_k$
  and $\gamma \in \spt{u}$ each term
  $|\trace{(\tilde{G_j} C_{j \gamma})} - (g_j \tilde{\sigma})_\gamma|$
  related to the constraint $g_j \geq 0$ as well as each term (omitted
  for conciseness) involving $\tilde{c}_\alpha$ related to the
  constraint $1 - X^{2 \alpha} \geq 0$.\\
  By using the fact that
  $\trace{(\tilde{G}_j C_{j\gamma})} = \sum_\delta g_{j \delta}
  \sum_{\alpha + \beta + \delta = \gamma} \tilde{G}_{j \alpha,\beta}$,
  we obtain
  \[
  |\trace{(\tilde{G_j} C_{j \gamma})} - (g_j \tilde{\sigma})_\gamma|
  \leq \sum_\delta |g_{j \delta}| \frac{\sqrt{r_j}(r_j+1)2^{-\delta_c}
    \, R}{1 - (r_j+1)2^{-\delta_c}} \,, \]
  where $r_j$ is the size of $\tilde{G_j}$.\\
  This yields an upper bound of
  $ d^n 2^{\tau+1} \frac{\sqrt{r_0}(r_0+1)2^{-\delta_c} \, R}{1 -
    (r_0+1)2^{-\delta_c}}$.
  We obtain a similar bound (omitted for conciseness) for each term
  involving $\tilde{c}_\alpha$.
  Then, we take the smallest $\delta$ such that
  $2^{-\delta} \leq \frac{\epsilon}{2 r_0}$ and the smallest
  $\delta_c$ such that
  \[ d^n 2^{\tau} \frac{\sqrt{r_0}(r_0+1)2^{-\delta_c} \, R}{1 -
    (r_0+1)2^{-\delta_c}} \leq \frac{\varepsilon}{2 r_0 ((m+1) +
    r_0)} \,. \]
  Thus, one can choose $\delta$ and $\delta_c$ of bit size upper bounded by 
\revise{$((m+1) D)^{D^{\bigo{(n)}} }$} in order to ensure that $\putinarsos$ terminates. As in the proof of Proposition~\eqref{th:bitmultivsos}, one shows that the  output is an exact Putinar's representation with rational coefficients of maximum bit
  size bounded by \revise{$((m+1) D^n)^{\bigo{(D^n)}}$}.
To bound the boolean run time, we
  consider the cost of solving SDP~\eqref{eq:dualsdp2}, which is
  performed in
  $\bigo{( \nsdp^4 \log_2(2^\tau \nsdp \, R \, 2^\delta) )}$
  iterations of the ellipsoid method, where each iteration requires
  $\bigo{(\nsdp^2(\msdp+\nsdp) )}$ arithmetic operations over
  $\log_2(2^\tau \nsdp \, R \, 2^\delta)$-bit numbers.  Since $\msdp$
  is bounded by $\binom{n+D}{n} \leq 2 D^n$, we obtain the desired result. 
\if{  
  and 
$\log_2 D = \bigo{(  2^{C_S \tau d^{2 n + 2}} )}$, 
one has  
  \[
  D^n = \bigo{\bigl( 2^{ n 2^{C_S \tau d^{2 n + 2}}  } \bigr)} \leq
  \bigo{\bigl( 2^{ 2^{(C_S + 1) \tau d^{2 n + 2}}  } \bigr)}
  \,.
  \]
  We obtain a similar bound for $\nsdp$, which ends the proof.
  }\fi
  }
\end{proof}

{The complexity is singly exponential in the degree $D$ of the
  representation. On all the examples we tackled, it was close to the degrees of
  the involved polynomials, as 
shown in Section~\ref{sec:benchs}.}

{
\subsection{Comparison with the rounding-projection algorithm of Peyrl and Parrilo}
\label{sec:ppPutinar}
We now state a constrained version of the rounding-projection algorithm from~\cite{PaPe08}. 
}
\begin{algorithm}
\caption{{$\PPcon$}%: algorithm to compute weighted SOS decompositions of multivariate polynomials in $\mathring{\Sigma}[X]$.
}
\label{alg:PPcon}
{\begin{algorithmic}[1]
%\Require $f \in \mathring{\Sigma}_\Z[X]$, positive $\varepsilon \in \Q$, precision parameters $\delta, R \in \N$ for the SDP solver 
\Require $f,g_1,\dots,g_m \in \Z[X]$ of degrees less than $d \in \N$, $\K := \{\x \in \R^n : g_1(\x) \geq 0, \dots, g_m(\x) \geq 0 \}$, rounding precision $\delta_i \in \N$, precision parameters $\delta, R \in \N$ for the SDP solver, precision $\delta_c \in \N$ for the Cholesky's decomposition
\Ensure lists $\clist_0,\dots,\clist_m$ of numbers in $\Q$ and lists $\slist_0,\dots,\slist_m$ of polynomials in $\Q[X]$
%\State $P := \polytope{(f)}$, $Q := P/2 \cap \N^n$\label{line:npPP}
%
\State $k \gets \lceil d / 2 \rceil$, $D \gets 2 k$, $g_0 := 1$
\While {$f  \notin \mathcal{Q}_D(\K)$} \label{line:PPDPuti} 
 $k \gets k + 1$, $D \gets D + 2$
\EndWhile \label{line:PPDPutf}
\State ok := false
\While {not ok} \label{line:PPicon}
\State $[\tilde{G_0},\dots,\tilde{G}_{m},  \tilde{\lambda}_0,\dots,\tilde{\lambda}_{m}], \gets \sdpconfun{f}{ \delta}{R}{\K}$ \label{line:sdpconPP}
\State $G' \gets \roundfun{\tilde{G}_0}{\delta_i}$
\For {$j \in \{1,\dots,m\}$}
\State $(s_{1 j},\dots,s_{r_j j}) \gets \choleskyfun{\tilde{G}_j}{\tilde{\lambda}_j}{\delta_c}$, $\tilde{\sigma}_j := \sum_{i=1}^{r_j} s_{i j}^2 $ \label{line:cholconPP}
\State $\clist_j \gets [1,\dots,1] $, $\slist_j \gets [s_{1 j},\dots,s_{r_j j}]$
\EndFor
\State $u \gets f - \sum_{j=1}^m \tilde{\sigma}_j$
\State $Q :=  \N_k^n$ \label{line:npPPcon}
\For {$\alpha, \beta \in Q$}
$\eta(\alpha+\beta) \gets \# \{(\alpha',\beta') \in Q^2 \mid \alpha'+\beta' = \alpha + \beta \}$
\State $G(\alpha,\beta) := G'(\alpha,\beta) - \frac{1}{\eta(\alpha+\beta)} \Bigl( \sum_{\alpha'+\beta'=\alpha + \beta} G'(\alpha',\beta')-u_{\alpha+\beta} \Bigr)$ \label{line:projPPcon}
\EndFor
\State $(c_{10},\dots, c_{r_0 0}, s_{1 0},\dots,s_{ r_0 0}) \gets \ldlfun{G}$ \label{line:ldlcon} \Comment{$f =   \sum_{i=1}^{r_0} c_{i 0}  s_{i 0}^2 + \sum_{j=1}^m \tilde{\sigma}_j$}
\If {$c_{10},\dots,c_{m r_m} \in \Q^{\geq 0}, s_{0 1},\dots,s_{m r_m} \in \Q[X]$} ok := true \label{line:ldlokcon}
\Else $\ \delta_i \gets 2 \delta_i$, $\ \delta \gets 2 \delta$, $R \gets 2 R$, $\delta_c \gets 2 \delta_c$
\EndIf
\EndWhile \label{line:PPfcon}
\State $\clist_0 \gets [c_{1 0},\dots,c_{r_0 0}] $, $\slist_0 \gets [s_{1 0},\dots,s_{r_0 0}]$
\State \Return $\clist_0,\dots,\clist_m, \slist_0,\dots,\slist_m$
\end{algorithmic}}
\end{algorithm}

{
For $f\in \Z[X]$ positive over a basic compact semi-algebraic set $S$ satisfying Assumption~\ref{hyp:arch}, Algorithm~$\PPcon$ starts as in Algorithm~$\putinarsos$ (see Section~\ref{sec:putinarsos}): it outputs the smallest $D$ such that $f \in \mathcal{Q}_D(S)$, solves SDP~\eqref{eq:dualsdp2} in Line~\lineref{line:sdpconPP}, and performs Cholesky's factorization in Line~\lineref{line:cholconPP}  to obtain an approximate Putinar's representation of $f$. Note that the approximate Cholesky's factorization is performed to obtain weighted SOS decompositions associated to the constraints $g_1,\dots, g_m$ (i.e.~$\tilde{\sigma}_1,\dots,\tilde{\sigma}_m$, respectively). \\
Next, the algorithm applies in Line~\lineref{line:projPPcon} the same projection procedure of Algorithm~$\PP$ (see Section~\ref{sec:pp}) on the polynomial $u := f - \sum_{j=1}^m \tilde{\sigma}_j g_j$. Note that when there are no constraints, one retrieves exactly the projection procedure from Algorithm~$\PP$. Exact $L D L^T$ is then performed on the matrix $G$ corresponding to $u$.\\
If all input precision parameters are large enough, $G$ is a Gram matrix associated to $u$ and $\tilde{\sigma}_1,\dots,\tilde{\sigma}_m$ are weighted SOS polynomals, yielding the exact Putinar's representation $f = u + \sum_{j=1}^m \tilde{\sigma}_j g_j$. As for Theorem~\ref{th:bitPP} and Theorem~\ref{th:putinarsos},  Algorithm~$\PPcon$  has a similar bit complexity than $\putinarsos$.
\if{
\begin{theorem}
\label{th:ppcon}
For some $C_\K > 0$ depending on $\K$, there exist $\delta_i$, $\varepsilon$,
$\delta$, $R$, $\delta_c$ and $D = 2 k$ of bit sizes less than 
$\bigo{(2^{\tau d^{n C_\K}})}$ for which
$\PPconfun{f}{S}{\delta_i}{\varepsilon}{\delta}{R}{\delta_c}$ terminates
and outputs an exact Putinar's representation with rational
coefficients of $f \in \mathcal{Q}(S)$. 
The maximum bit size of these coefficients is bounded by
$\bigo{(2^{\tau d^{n C_\K}})}$ and the procedure runs in boolean time
$\bigo{\bigl(2^{2^{\tau d^{n C_\K}}} \bigr)}$.
\end{theorem}
\begin{proof}
The bit complexity is analyzed as for Algorithm~$\PP$ (Theorem~\ref{th:bitPP}) and Algorithm~$\putinarsos$ (Theorem~\ref{th:putinarsos}).
\end{proof}
}\fi
}
%%% Local Variables:
%%% mode: latex
%%% TeX-master: "multivsos"
%%% End:

\section{Practical experiments}
\label{sec:benchs}
We provide experimental results for Algorithms~$\intsos$, $\reznicksos$
and $\putinarsos$. These are implemented in a procedure, called $\multivsos$,
and integrated in the $\realcertify$ library {by~\cite{RealCertify}}, written
  in Maple. More details about installation and benchmark execution are given on
  the dedicated
  webpage\footnote{\url{https://gricad-gitlab.univ-grenoble-alpes.fr/magronv/RealCertify}}.
%to univariate\footnote{\url{https://github.com/magronv/univsos}} and multivariate\footnote{\url{https://github.com/magronv/multivsos}} polynomials.  
%This tool is available within the RAGlib Maple package\footnote{\url{http://www-polsys.lip6.fr/~safey/RAGLib/}}.  
%
All results were obtained on an Intel Core i7-5600U CPU (2.60 GHz) with 16Gb of
RAM. We use the Maple~\texttt{Convex}
package\footnote{\url{http://www.home.math.uwo.ca/faculty/franz/convex}} to
compute Newton polytopes. Our subroutine $\sdp$ relies on the
arbitrary-precision solver SDPA-GMP by~\cite{Nakata10GMP} and the $\cholesky$
procedure is implemented with  \texttt{LUDecomposition} available
within Maple. Most of the time is spent in the $\sdp$ procedure for all
benchmarks. {To decide non-negativity of polynomials, we use either
  $\raglib$ or the $\sdp$ procedure as oracles. {Recall that $\raglib$ relies on
  critical point methods whose runtime strongly depends on the number of
  (complex) solutions to polynomial systems encoding critical points. While
  these methods are more versatile, this number is
  generically exponential in $n$. Hence, we prefer to rely at first on a {\em
    heuristic} strategy based on using $\sdp$ first (recall that it does not
  provide an exact answer).}  }

\begin{table}[!t]
\begin{center}
\caption{$\multivsos$ vs $\univsostwo$ \cite{univsos} for benchmarks from~\cite{Chevillard11}.}
\begin{tabular}{lcr|rr|rr}
\hline
\multirow{2}{*}{Id} & \multirow{2}{*}{$d$} & \multirow{2}{*}{$\tau$ (bits)} & \multicolumn{2}{c|}{$\multivsos$} & \multicolumn{2}{c}{$\univsostwo$} \\
 & & & $\tau_1$ (bits) & $t_1$ (s)  & $\tau_2$ (bits) & $t_2$ (s)  \\
\hline  
\# 1 & 13 & 22 682 & 387 178 & 0.84 & 51 992 & 0.83  \\
\# 3 & 32 & 269 958 & $-$ & $-$ & 580 335 & 2.64  \\
\# 4 & 22 & 47 019 &  1 229 036 & 2.08  & 106 797 & 1.78 \\
\# 5 & 34 & 117 307 & 10 271 899 & 69.3  &   265 330 & 5.21 \\
\# 6 & 17 & 26 438 & 713 865 & 1.15  & 59 926 & 1.03 \\
\# 7 & 43 & 67 399 & 10 360 440 &  16.3  & 152 277 & 11.2  \\
\# 8 & 22 & 27 581 & 1 123 152 & 1.95 & 63 630 & 1.86 \\
\# 9 & 20 & 30 414 & 896 342 & 1.54  & 68 664 & 1.61 \\
\# 10 & 25 & 42 749 & 2 436 703 & 3.02  &  98 926 & 2.76  \\
\hline
\end{tabular}
\label{table:bench1}
\end{center}
\end{table} 

In Table~\ref{table:bench1}, we compare the performance of $\multivsos$ for nine
univariate polynomials being positive over compact intervals. More details about
these benchmarks are given in~\cite[Section~6]{Chevillard11}
and~\cite[Section~5]{univsos}. In this case, we use $\putinarsos$. The main
difference is that we use SDP in $\multivsos$ instead of complex root isolation
in $\univsostwo$. The results emphasize that $\univsostwo$ is faster and
provides more concise SOS certificates, especially for high degrees (see e.g.~\#
5). For \# 3, we were not able to obtain a decomposition within a day of
computation with $\multivsos$, as meant by the symbol $-$ in the corresponding
column entries. Large values of $d$ and $\tau$ require more precision. The
values of $\varepsilon$, $\delta$ and $\delta_c$ are respectively between
$2^{-80}$ and $2^{-240}$, 30 and 100, 200 and 2000.

\begin{table}[!ht]
\begin{center}
\caption{ $\multivsos$ vs $\PP$~\cite{PaPe08} vs $\raglib$ vs $\cad$ (Reznick).} 
% for $n$-variate polynomials of degree $d$
%
\begin{tabular}{lrr|rrrr|c|c}
\hline
\multirow{2}{*}{Id} & \multirow{2}{*}{$n$} & \multirow{2}{*}{$d$}  & \multicolumn{2}{c}{$\multivsos$} & \multicolumn{2}{c|}{$\PP$} & $\raglib$ & $\cad$ \\
 & & & $\tau_1$ (bits) & $t_1$ (s)  & $\tau_2$ (bits) & $t_2$ (s) & $t_3$ (s) & $t_4$ (s) \\
\hline  
$f_{12}$ & 2 & 12 & 316 479 & 3.99 & 3 274 148 & 3.87 & 0.15 & 0.07 \\
$f_{20}$ & 2 & 20 & 754 168 & 113. & 53 661 174 & 137. & 0.16 & 0.03 \\
$M_{20}$ & 3 & 8 & 4 397 & 0.14 & 3 996 & 0.16 & 0.13 & 0.05 \\
$M_{100}$ & 3 & 8 & 56 261 & 0.26 & 12 200 & 0.20 & 0.15 & 0.03 \\
$r_2$ & 2 & 4 & 1 680 & 0.11 & 1 031 & 0.12 & 0.09 & 0.01 \\
$r_4$ & 4 & 4 & 13 351 & 0.14 & 47 133  & 0.15 & 0.32 & $-$ \\
$r_6$ & 6 & 4 & 52 446 & 0.24 & 475 359 & 0.37 & 623. & $-$ \\
$r_8$ & 8 & 4 & 145 933 & 0.70 & 2 251 511 & 1.08 & $-$ & $-$ \\
$r_{10}$ & 10 & 4 & 317 906 & 3.38 & 8 374 082 & 4.32 & $-$ & $-$ \\
$r_6^2$ & 6 & 8 & 1 180 699 & 13.4 & 146 103 466 & 112. & 10.9 & $-$ \\
\hline
\end{tabular}
\label{table:bench2}
\end{center}
\end{table} 

Next, we compare the performance of $\multivsos$ with other tools in
Table~\ref{table:bench2}. The two first benchmarks are built from the
polynomial
$f = (X_1^2+1)^2+(X_2^2+1)^2 + 2 (X_1+X_2+1)^2 - 268849736/10^8$
from~\cite[Example~1]{Las01sos}, with $f_{12} := f^3$ and
$f_{20} := f^5$. For these two benchmarks, we apply $\intsos$. We use
$\reznicksos$ to handle $M_{20}$ (resp.~$M_{100}$), obtained as in
Example~\ref{ex:reznick} by adding $2^{-20}$ (resp.~$2^{-100}$) to the
positive coefficients of the Motzkin polynomial and $r_{i}$, which is
a randomly generated positive definite quartic with $i$ variables. We
implemented in Maple the projection and rounding algorithm
from~\cite{PaPe08} {(stated in Section~\ref{sec:pp})} also relying on SDP, denoted by~$\PP$. For
$\multivsos$, the values of $\varepsilon$, $\delta$ and $\delta_c$ lie
between $2^{-100}$ and $2^{-10}$, 60 and 200, 10 and 60.

{ In most cases, $\multivsos$ is more efficient than $\PP$
  and outputs more concise representations. The reason is that $\multivsos$
  performs approximate Cholesky's decompositions while $\PP$ computes exact $L D
  L^T$ decompositions of Gram matrices obtained after the two steps of rounding
  and projection. 
%  This observation matches with the theoretical complexity   estimates established in Proposition~\ref{th:bitmultivsos} and  Theorem~\ref{th:bitPP}. 
  Note that we could not solve the examples of
  Table~\ref{table:bench2} with less precision.
}

We compare with~$\raglib$~\cite{raglib} based on critical point methods~(see
e.g. \cite{SaSc03, HS12}) and the~\texttt{SamplePoints}
procedure~\cite{Lemaire05} (abbreviated as~$\cad$) based on
CAD~\cite{Collins75}, both available in Maple. Observe that $\multivsos$ can
tackle examples which have large degree but a rather small number of variables
($n \leq 3$) and then return certificates of non-negativity. The runtimes are
slower than what can be obtained with  $\raglib$ and/or
$\cad$ (which in this setting have polynomial complexity when $n\leq 3$ is
fixed). Note that the bit size of the certificates which are obtained here is
quite large which explains this phenomenon.  

% These methods outperform the two SDP-based algorithms for examples with $n \leq
% 3$ and note that $\raglib$ can handle many more problems than CAD based
% implementations. However, they are both slower for examples such as $r_6^2$ when
% the number of variables increases.

When the number of variables increases, $\cad$ cannot reach many of the problems
we considered. Note that $\multivsos$ becomes not only faster but can solve
problems which are not tractable by $\raglib$. 

{Recall that $\multivsos$ relies first on solving numerically Linear
  Matrix Inequalities ; this is done at finite precision in time polynomial in
  the size of the input matrix, which, here is bounded by $\binom{n+d}{d}$.
  Hence, at fixed degree, that quantity evolves polynomially in $n$. On the
  other hand, the quantity which governs the behaviour of fast implementations
  based on the critical point method is the degree of the critical locus of some
  map. On the examples considered, this degree matches the worst case bound
  which is the B\'ezout number $d^n$. Besides, the doubly exponential
  theoretically proven complexity of CAD is also met on these examples. }

{These examples illustrate the potential of $\multivsos$ and more
  generally SDP-based methods: at fixed degree, one can hope to take advantage
  of fast numerical algorithms for SDP and tackle examples involving more
  variables than what could be achieved with more general tools. }

{Recall however that $\multivsos$ computes rational certificates of
  non-negativity in some ``easy'' situations: roughly speaking, these are the
  situations where the input polynomial lies in the interior of the SOS cone and
  has coefficients of moderate bit size. This fact is illustrated by
  Table~\ref{table:bench4}.}

\begin{table}[!ht]
{
\begin{center}
\caption{$\multivsos$ vs $\raglib$ vs $\cad$ for non-negative polynomials which are presumably not in $\mathring{\Sigma}[X]$.}
\begin{tabular}{lrr|cc|r|c}
\hline
\multirow{2}{*}{Id} & \multirow{2}{*}{$n$} & \multirow{2}{*}{$d$} & \multicolumn{2}{c|}{$\multivsos$} & $\raglib$ & $\cad$ \\
 & & & $\tau_1$ (bits) & $t_1$ (s) & $t_2$ (s) & $t_3$ (s)\\
  \hline
  $S_{1}$ & 4 & 24 & $-$  & $-$  & 1788.  &$-$ \\
  $S_{2}$ & 4 & 24 & $-$ & $-$   & 1840.  &$-$ \\
  $V_{1}$ & 6 & 8 & $-$ & $-$   & 5.00  &$-$ \\
  $V_{2}$ & 5 & 18 & $-$ & $-$   & 1180.  &$-$ \\
  $M_{1}$ & 8 & 8 & $-$ & $-$  & 351.  &$-$\\
  $M_{2}$ & 8 & 8 & $-$ & $-$  & 82.0  &$-$\\
  $M_{3}$  & 8 & 8 & $-$ & $-$  & 120.  &$-$\\
  $M_{4}$ & 8 & 8 & $-$ & $-$  & 84.0  & $-$\\
\hline
\end{tabular}
\label{table:bench4}
\end{center}
}
\end{table}

{ This table reports on problems appearing enumerative geometry
  (polynomials $S_1$ and $S_2$ communicated by Sottile and appearing in the
  proof of the Shapiro conjecture \cite{Sot00}), computational geometry
  (polynomials $V_1$ and $V_2$ appear in \cite{ELLS07}) and in the proof of the
  monotone permanent conjecture in~\cite{Haglund1999} ($M_1$ to $M_4$). }

{We were not able to compute certificates of non-negativity for these
  problems which we presume do not lie in the interior of the SOS cone. This
  illustrates the current theoretical limitation of $\multivsos$. These problems
  are too large for $\cad$ but $\raglib$ can handle them. Note that some of
  these examples involve $8$ variables ; we observed that the B\'ezout number is
  far above the degree of the critical loci computed by the critical point
  algorithms in $\raglib$. This explains the efficiency of such tools on these
  problems. Recall however that $\raglib$ did not provide a certificate of
  non-negativity.}

{This whole set of examples illustrates first the efficiency and
  usability of $\multivsos$ as well as its complementarity with other more
  general and versatile methods. This shows also the need of further research to handle in a
  systematic way more general non-negative polynomials than what it does
  currently. For instance, we emphasize that certificates of non-negativity were
  computed for $M_i$ ($1\leq i \leq 4$) in~\cite{Kaltofen09} (see also
  \cite{KLYZ08}).}

\begin{table}[!ht]
{
\begin{center}
\caption{$\multivsos$ vs $\PPcon$ vs $\raglib$ vs $\cad$ (Putinar).}
\begin{tabular}{lrr|rrrrr|c|c}
\hline
\multirow{2}{*}{Id} & \multirow{2}{*}{$n$} & \multirow{2}{*}{$d$}  & & \multicolumn{2}{c}{$\multivsos$} & \multicolumn{2}{c|}{$\PP$} & $\raglib$ & $\cad$ \\
 & & & $k$ & $\tau_1$ (bits) & $t_1$ (s) & $\tau_2$ (bits) & $t_2$ (s) & $t_3$ (s) & $t_4$ (s) \\
\hline  
$p_{46}$ & 2 & 4 & 3 & 45 168 & 0.17 & 230 101 & 0.19 & 0.15 & 0.81 \\
$f_{260}$ & 6 & 3 & 2 & 251 411 & 2.35 & 5 070 043 & 3.60 & 0.12 & $-$ \\
$f_{491}$ & 6 & 3 & 2 & 245 392 & 4.63 & 4 949 017 & 5.63 & 0.01 & 0.05 \\
$f_{752}$ & 6 & 2 & 2 & 23 311 & 0.16 & 74 536 & 0.15 & 0.07 & $-$ \\
$f_{859}$ & 6 & 7 & 4 & 13 596 376 & 299. & 2 115 870 194 & 5339. & 5896. & $-$ \\
$f_{863}$ & 4 & 2 & 1 & 12 753 & 0.13 & 30 470 & 0.13 & 0.01 & 0.01 \\
$f_{884}$ & 4 & 4 & 3 & 423 325 & 13.7 & 10 122 450 & 16.1 & 0.21 & $-$ \\
$f_{890}$ & 4 & 4 & 2 & 80 587 & 0.48 & 775 547 & 0.56 & 0.08 & $-$ \\
butcher & 6 & 3 & 2 & 538 184 & 1.36 & 8 963 044 & 3.35 & 47.2 & $-$ \\
heart & 8 & 4 & 2 & 1 316 128 & 3.65 & 35 919 125 & 14.1 & 0.54 & $-$ \\
magnetism & 7 & 2 & 1 & 19 606 & 0.29 & 16 022 & 0.28 & 434. & $-$ \\
\hline
\end{tabular}
\label{table:bench3}
\end{center}
}
\end{table}

Finally, we compare the performance of $\multivsos$ ($\putinarsos$) on
positive polynomials over basic compact semi-algebraic sets in
Table~\ref{table:bench3}. 
The first benchmark is
from~\cite[Problem~4.6]{Las01sos}. Each benchmark $f_i$ comes from an
inequality of the Flyspeck project~\cite{Hales_theflyspeck}. The three
last benchmarks are from~\cite{Munoz13}. The maximal degree of the
polynomials involved in each system is denoted by $d$. We emphasize
that the degree $D = 2k$ of each Putinar representation obtained in
practice with $\putinarsos$ is very close to $d$, which is in contrast
with the theoretical complexity estimates obtained in
Section~\ref{sec:putinar}. The values of $\varepsilon$, $\delta$ and
$\delta_c$ lie between $2^{-30}$ and $2^{-10}$, 60 and 200, 10 and 30.\\
As for Table~\ref{table:bench2}, $\raglib$ and $\multivsos$ can both solve large
problems (involving e.g. $8$ variables) but note that $\multivsos$ outputs
certificates of emptiness which cannot be computed with implementations based on
the critical point method. In terms of timings, $\multivsos$ is sometimes way
faster (e.g.~magnetism, $f_{859}$) but that it is hard here to draw some general
rules. Again, it is important to keep in mind the parameters which influence the
runtimes of both techniques. As before, for $\multivsos$, the size of the SDP to
be solved is clearly the key quantity. Also, it is important to write the
systems in an appropriate way also to limit the size of those matrices (e.g.
write $1-x^2\leq 0$ to model $-1\leq x \leq 1$). For $\raglib$, it is way better
to write $-1\leq x$ and $x\leq 1$ to better control the B\'ezout bounds
governing the difficulty of solving systems with purely algebraic methods. Note
also that the number of inequalities increase the combinatorial complexity of
those techniques. 

Finally, note that $\cad$ can only solve 3 benchmarks out of 10 and all in all
$\multivsos$ and $\raglib$ solve a similar amount of problems; the latter one
however does not provide certificates of emptiness. As for
Table~\ref{table:bench2}, $\multivsos$ and
$\PPcon$ yield similar performance, while the former provides more concise
output than the latter.

%%% Local Variables:
%%% mode: latex
%%% TeX-master: "multivsos"
%%% End:

\section{Conclusion and perspectives}
{
We designed and analyzed new algorithms to compute rational SOS decompositions for several sub-classes of non-negative multivariate polynomials, including  positive definite forms and polynomials positive over basic compact semi-algebraic sets.
Our framework relies on SDP solvers implemented with interior-point methods. A drawback of such methods, in the context of unconstrained polynomial optimization, is that we are restricted to non-negative polynomials belonging to the interior of the SOS cone. 
We shall investigate the design of specific algorithms for the sub-class of polynomials lying in the border of the SOS cone. 
We also plan to adapt our framework, either for problems involving non-commutative polynomial data, or for alternative certification schemes, e.g. in the context of linear/geometric programming relaxations.
}

\appendix
\if{
% -*- coding: utf-8 -*-
%!TEX encoding = UTF-8 Unicode
% Created 2015-06-16 Tue 13:52
\documentclass[a4paper,10pt]{article}
\textheight235mm
\textwidth160mm
\voffset-10mm
\hoffset-10mm
\parindent0cm
\parskip2mm

\usepackage{amsmath}
% Copyright
%\setcopyright{none}
%\setcopyright{acmcopyright}
%\setcopyright{acmlicensed}
%\setcopyright{rightsretained}
%\setcopyright{usgov}
%\setcopyright{usgovmixed}
%\setcopyright{cagov}
%\setcopyright{cagovmixed}

% DOI
%\acmDOI{10.475/123_4}

% ISBN
%\acmISBN{123-4567-24-567/08/06}

%Conference
%\acmConference[ISSAC'18]{ACM ISSAC conference}{July 2018}{New-York, USA} 
%\acmYear{2018}
%\copyrightyear{2018}

%\acmPrice{15.00}

%\acmSubmissionID{123-A12-B3}

%\usepackage[utf8]{inputenc}
%\usepackage[T1]{fontenc}
%\usepackage{fixltx2e}
\usepackage[utf8]{inputenc}
\usepackage[T1]{fontenc}
\usepackage{graphicx}
\usepackage{hyperref}
\usepackage{longtable}
\usepackage{multirow}
\usepackage{float}
\usepackage{wrapfig}
\usepackage{rotating}
\usepackage[normalem]{ulem}
\usepackage{amsmath}
\usepackage{amsthm}
\usepackage{textcomp}
\usepackage{marvosym}
\usepackage{wasysym}
\usepackage{amssymb}
\usepackage{capt-of}
\usepackage{bm}
% \usepackage{natbib}
% \usepackage[pdftex,                %
%     bookmarks         = true,%     % Signets
%     bookmarksnumbered = true,%     % Signets numérotés
% %    pdfpagemode       = None,%     % Signets/vignettes fermé à l'ouverture
%     pdfstartview      = FitH,%     % La page prend toute la largeur
%     pdfpagelayout     = SinglePage,% Vue par page
%     colorlinks        = true,%     % Liens en couleur
%     linkcolor= blue, %    % couleur des liens internes
%     anchorcolor= blue, %    % couleur des liens internes
%     citecolor         =blue,
%     urlcolor          = magenta,%  % Couleur des liens externes
% %    pdfborder         = {0 0 0}%   % Style de bordure : ici, pas de bordure
%     ]{hyperref}%                   % Utilisation de HyperTeX

\usepackage{comment}
\usepackage{color}
\usepackage{enumerate} 
\usepackage{myalgo}%markus
\usepackage{bm}

\newcommand{\coq}{\text{\sc Coq}}
\newcommand{\hol}{\text{\sc Hol-light}}

\newcommand{\mohab}[1]{{\color{green} Mohab: #1}}
\newcommand{\victor}[1]{{\color{red} Victor: #1}}

\newcommand{\newjsc}[1]{\textbf{{\color{blue}#1}}}
\tolerance=1000

\providecommand{\alert}[1]{\textbf{#1}}
\def\A{\mathbf{A}}
\def\B{\mathbf{B}}
\def\I{\mathbf{I}}
\def\Sb{\mathbb{S}}
\def\K{S}
\def\X{\mathbf{X}}
\def\Y{\mathbf{Y}}
\def\bmx{\bm{x}}
\def\x{\bmx}
\def\bmy{\bm{y}}
\def\y{\bmy}
\def\p{\mathbf{p}}
\def\f{\mathbf{f}}
\def\q{\mathbf{q}}
\def\z{\mathbf{z}}
\def\M{\mathbf{M}}

\DeclareMathOperator{\polytope}{\mathcal{C}}

\newcommand{\R}{\mathbb{R}}
\newcommand{\Z}{\mathbb{Z}}
\newcommand{\Q}{\mathbb{Q}}
\newcommand{\N}{\mathbb{N}}
\newcommand{\red}[1]{\textbf{{\color{red}#1}}}
\DeclareMathOperator{\bigo}{\mathcal{O}}
\DeclareMathOperator{\sgn}{sgn}
\DeclareMathOperator{\trace}{Tr}
\DeclareMathOperator{\bigotilde}{\overset{\sim}{\mathcal{O}}}
\newcommand{\hasrealroots}{\texttt{has\_real\_roots}}
\newcommand{\hasrealrootsfun}[1]{\texttt{has\_real\_roots}(#1)}
\newcommand{\newtonpolytope}{\texttt{newton\_polytope}}
\newcommand{\newtonpolytopefun}[1]{\texttt{newton\_polytope}(#1)}
\newcommand{\sumtwosquares}{\texttt{sum\_two\_squares}}
\newcommand{\sumtwosquaresfun}[2]{\texttt{sum\_two\_squares}(#1,#2)}
\newcommand{\sumofsquares}{\texttt{sos}}
\newcommand{\sumofsquaresfun}[2]{\texttt{sos}(#1,#2)}
\newcommand{\sdp}{\texttt{sdp}}
\newcommand{\sdpcon}{\texttt{sdp}}
\newcommand{\sdpHA}{\texttt{sdp}}
\newcommand{\roundfun}[2]{\texttt{round}(#1,#2)}
\newcommand{\ldlfun}[1]{\texttt{ldl}(#1)}
\newcommand{\sdpfun}[3]{\texttt{sdp}(#1,#2,#3)}
\newcommand{\sdpfunHA}[4]{\texttt{sdp}(#1,#2,#3,#4)}
\newcommand{\sdpconfun}[4]{\texttt{sdp}(#1,#2,#3,#4)}
\newcommand{\cholesky}{\texttt{cholesky}}
\newcommand{\choleskyfun}[3]{\texttt{\cholesky}(#1,#2,#3)}
\newcommand{\soslist}{\texttt{sos\_list}}
\newcommand{\nsdp}{n_\text{sdp}}
\newcommand{\msdp}{m_\text{sdp}}
\newcommand{\qlist}{\texttt{q\_list}}
\newcommand{\hlist}{\texttt{h\_list}}
\newcommand{\clist}{\texttt{c\_list}}
\newcommand{\calpha}{\texttt{c\_alpha}}
\newcommand{\slist}{\texttt{s\_list}}
\newcommand{\univsos}{\texttt{univsos}}
\newcommand{\univsosone}{\texttt{univsos1}}
\newcommand{\univsostwo}{\texttt{univsos2}}
\newcommand{\intsos}{\texttt{intsos}}
\newcommand{\multivsos}{\texttt{multivsos}}
\newcommand{\realcertify}{\texttt{RealCertify}}
\newcommand{\intsosfun}[5]{\texttt{intsos}(#1,#2,#3,#4,#5)}
\newcommand{\absorb}{\texttt{absorb}}
\newcommand{\absorbfun}[5]{\texttt{absorb}(#1,#2,#3,#4,#5)}
\newcommand{\putinarsosfun}[6]{\texttt{Putinarsos}(#1,#2,#3,#4,#5,#6)}
\newcommand{\polyasos}{\texttt{Polyasos}}
\newcommand{\hilbertsos}{\texttt{Hilbertsos}}
\newcommand{\hilbertsosfun}[5]{\texttt{Hilbertsos}(#1,#2,#3,#4,#5)}
\newcommand{\cad}{\texttt{CAD}}
\newcommand{\PP}{\texttt{RoundProject}}
\newcommand{\PPfun}[4]{\texttt{RoundProject}(#1,#2,#3,#4)}
\newcommand{\PPcon}{\texttt{RoundProjectPutinar}}
\newcommand{\raglib}{\texttt{RAGLib}}
\newcommand{\putinarsos}{\texttt{Putinarsos}}
\newcommand{\multivsosone}{\texttt{multivsos1}}
\newcommand{\multivsostwo}{\texttt{multivsos2}}

\theoremstyle{plain}
\newtheorem{theorem}{Theorem}[section]
\newtheorem{lemma}[theorem]{Lemma}
\newtheorem{proposition}[theorem]{Proposition}
\newtheorem{corollary}[theorem]{Corollary}

\theoremstyle{definition}
\newtheorem{definition}[theorem]{Definition}
\newtheorem{assumption}[theorem]{Assumption}
\newtheorem{example}{Example}
\newtheorem{remark}{Remark}

%\theoremstyle{plain}
%\newtheorem{assumption}{Assumption}

% \theoremstyle{plain}
% \newtheorem{theorem}{Theorem}[section]
% \newtheorem{lemma}[theorem]{Lemma}
% \newtheorem{proposition}[theorem]{Proposition}
% \newtheorem{corollary}[theorem]{Corollary}
% \newtheorem{question}{Question}

% \theoremstyle{definition}
% \newtheorem{definition}[theorem]{Definition}
% %\newtheorem{hypothesis}[theorem]{Assumption}
% %\newtheorem{conjecture}[theorem]{Conjecture}
%\newtheorem{remark}{Remark}
% \newtheorem{assumption}[theorem]{Assumption}
% \newtheorem{example}{Example}

%\newenvironment{proof}[1][Proof]{\begin{trivlist}
%\item[\hskip \labelsep {\bfseries #1}]}{\end{trivlist}}
%\newenvironment{definition}[1][Definition]{\begin{trivlist}
%\item[\hskip \labelsep {\bfseries #1}]}{\end{trivlist}}
%\newenvironment{example}[1][Example]{\begin{trivlist}
%\item[\hskip \labelsep {\bfseries #1}]}{\end{trivlist}}
%\newenvironment{remark}[1][Remark]{\begin{trivlist}
%\item[\hskip \labelsep {\bfseries #1}]}{\end{trivlist}}

\if{
\newcommand{\qed}{\nobreak \ifvmode \relax \else
      \ifdim\lastskip<1.5em \hskip-\lastskip
      \hskip1.5em plus0em minus0.5em \fi \nobreak
      \vrule height0.75em width0.5em depth0.25em\fi}
}\fi
\newcommand{\anon}[0]{\,\centerdot\,}
\newcommand{\spt}[1]{\mbox{supp}(#1)}
\newcommand{\bracket}[2]{\langle #1,#2\rangle}

\DeclareMathOperator{\vol}{vol}

\def\mohab#1{\textcolor{magenta}{#1}}
\definecolor{dkviolet}{rgb}{0.6,0,0.8}
\def\victor#1{\textcolor{dkviolet}{#1}}
\parindent0cm

\begin{document}
\title{...}
\author{Victor Magron$^{1,2}$ \and Mohab Safey El Din$^{3}$}
\date{\today}

\maketitle

\footnotetext[1]{CNRS, L2S CENTRALESUPELEC, 3 Rue Joliot-Curie,  91192 Gif sur Yvette, France.}
\footnotetext[2]{CNRS LAAS, Universit\'e de Toulouse, 7 avenue du colonel Roche, F-31400 Toulouse, France.}
\footnotetext[3]{Sorbonne Universit\'e, CNRS, INRIA, Laboratoire d'Informatique de Paris 6, PolSys, Paris, France.}
%\footnotetext[3]{Laboratoire d'Informatique de Paris~6, LIP6, \'Equipe \textsc{PolSys}, 4 place Jussieu F-75252, Paris Cedex 05, France}
%\footnotetext[4]{Mohab Safey El Din is supported by the ANR grant {\sc Games} and the PGMO grant {\sc Gamma}.}

%\email{victor.magron@polsys.lip6.fr}
%\email{mohab.safey@lip6.fr}

%\author{Victor Magron$^{1,2}$ \and Mohab Safey El Din$^{2}$}
% \footnotetext[1]{CNRS Verimag; 700 av Centrale 38401 Saint-Martin d'H\`eres, France} 
% \footnotetext[2]{Sorbonne Universit\'es, UPMC Univ. Paris 06, CNRS, Inria Paris Center, LIP6, Equipe PolSys, F-75005, Paris, France}

\def\xxx{{\bf (xxx)}}
\def\todo{{\bf (todo!)}}
\def\check{{\bf (check!)}}

}\fi

\section{Appendix}
{
Let $f \in \Z[X_1, \ldots, X_n]$ of degree $d$ and $\tau$ be the maximum bit
size of the coefficients of $f$ in the standard monomial basis.
}

{
Let $V\subset \mathbb{C}^n$ be the algebraic set defined by
\begin{equation}\label{eq:1}
f = \frac{\partial f}{\partial X_2} = \cdots = \frac{\partial f}{\partial X_n} = 0
\end{equation}
By the algebraic version of Sard's theorem (see e.g.~\cite[Appendix B]{SaSc17}), when $V$ is
equidimensional and has at most finitely singular points, the projection of the
set $V\cap \R^n$ on the $X_1$-axis is finite (and hence a real algebraic set of
$\R$); we denote it by $Z_\R$. Hence, it is defined by the vanishing of some
polynomial in $\Z[X_1]$.

The goal of this Appendix is to provide a proof of
Proposition~\ref{prop:ori-cpm} which states that 
%\begin{proposition}
%  \label{prop:ori-cpm}
{under the above notation and assumption, there exists a polynomial $w\in \Z[X_1]$ of degree
$\leq d^n$ with coefficients of bit size $\leq \tau \cdot (4 d+2)^{3n}$ such that its set of
real roots contains $Z_\R$.} 
%\end{proposition}
%
To prove Proposition~\ref{prop:ori-cpm}, our strategy is to rely on
algorithms computing sample points in real algebraic sets: letting $C\subset V$
be a finite set of points which meet all connected components of $V\cap\R^n$, it
is immediate that the projection of $C$ on the $X_1$-axis contains $Z_\R$.\\
From the computation of an exact representation of such a set $C$, one will be
able to analyze the bit size of a polynomimal whose set of roots contains
$Z_\R$. We focus on algorithms based on the critical point method. Those yield
the best complexity estimates which are known in theory and practical
implementations reflecting these complexity gains have been obtained in
\cite{raglib} from e.g. \cite{SaSc03, HS12}. Here, we focus on \cite[Algorithm 13.3]{BPR06} since
it is the more general one and it does not depend on probabilistic choices which
make it easy to analyze from a bit complexity perspective.\\
It starts by computing the polynomial
\[
g = f^2 + \left (\frac{\partial f}{\partial X_2}\right )^2 + \cdots +\left (
\frac{\partial f}{\partial X_n}\right )^2.
\]
Observe that the set of real solutions of $g=0$ coincides with $V\cap \R^n$.
Next, one introduces two infinitesimals $\epsilon$ and $\eta$ (see \cite[Chap.
2]{BPR06} for an introduction on Puiseux series and infinitesimals). Consider
the polynomial:
\[
  g_1  = g + \left (\eta (X_1^2+\cdots + X_{n+1}^2) - 1\right )^2. 
\]
Its vanishing set over $\R\langle \eta \rangle^{n+1}$ corresponds to the
intersection of the lifting of the vanishing set of $g$ in $\R^n$ with the
hyperball of $\R\langle \eta \rangle^{n+1}$ centered at the origin of radius
$\frac{1}{\eta}$.\\
Let $d_i$ be the degree of $g_1$ in $X_i$. Without loss of generality, up to
reordering the variables, we assume that $d_1\geq d_2 \geq \cdots \geq d_n$ ; we
assume that after this process $X_1$ has been sent to $X_k$. Now, we let
\begin{align*}
  h & = g_1 (1-\varepsilon) + \varepsilon(X_1^{2(d_1+1)}+\cdots+X_n^{2(d_n+1)}+X_{n+1}^6 - (n+1)\zeta^{d+1})
\end{align*}
We finally focus on the polynomial system:
\[
h = \frac{\partial h}{\partial X_2}= \cdots = \frac{\partial h}{\partial
  X_{n+1}} = 0
\]
The rationale behind the last infinitesimal deformation is twofold (see
\cite[Chap. 12 and Chap. 13]{BPR06}):
\begin{itemize}
\item the algebraic set defined by the vanishing of $h$ is smooth ;
\item the above polynomial system is finite and forms a Gr\"obner basis $G$ for
  any degree lexicographical ordering with $X_1\succ \cdots\succ X_{n+1}$.
\end{itemize}
Besides, \cite[Prop. 13.30]{BPR06} states that taking the limits (when
infinitesimals tend to zero) of projections on the $(X_1, \ldots, X_n)$-space of
a finite set of points meeting each connected component of the real algebraic
set defined by $h = 0$ provides a finite set of points in the real algebraic set
defined by $g = 0$.\\
In our situation, we do not need to to go into such details. We only need to
compute a non-zero polynomial $w\in \Z[X_k]$ whose set of real roots contains
$Z_\R$. Using Stickelberger's theorem \cite[Theorem 4.98]{BPR06} and the process for computing
limits in~\cite[Algoroithm 12.14]{BPR06} and~\cite{RRS}, it suffices to compute the
characteristic polynomial of the multiplication operator by $X_k$ in the ring of
polynomials with coefficients in $\Q[\eta, \zeta]$ quotiented by the ideal
$\langle G \rangle$. This is done using \cite[Algorithm 12.9]{BPR06}.\\
In order to analyze the bit size of the coefficients of the output
characteristic polynomial, we need to bound the bit size of the entries in the
matrix output by \cite[Algorithm 12.9]{BPR06}. Following the discussion in the
complexity analysis of \cite[Algorithm 13.1]{BPR06}, we deduce that the
coefficients of these entries have bit size dominated by $\tau \left
  (2(2d+1)\right )^{2n}$. Besides, this matrix has size bounded by $\left
  (2(2d+1)\right )^{2n}$. We deduce that the coefficients of its characteristic
polynomial have bit size bounded by $\left (2(2d+1)\right )^{3n}$.
}
%\bibliographystyle{plain}
%\bibliography{multivsos}

%\end{document}

%\bibliographystyle{elsarticle-harv}
%\bibliography{multivsos}
% Include the ".bib" file (generated by bibtex) right here.

\end{document}